\documentclass[%
 aip,
 amsmath,amssymb,
 reprint,%
]{revtex4-1}

\usepackage{graphicx}
\usepackage{dcolumn}
\usepackage{bm}

\usepackage[utf8]{inputenc}
\usepackage[T1]{fontenc}
\usepackage{mathptmx}
\usepackage{etoolbox}

\makeatletter
\def\@email#1#2{%
 \endgroup
 \patchcmd{\titleblock@produce}
  {\frontmatter@RRAPformat}
  {\frontmatter@RRAPformat{\produce@RRAP{*#1\href{mailto:#2}{#2}}}\frontmatter@RRAPformat}
  {}{}
}%
\makeatother
\begin{document}

\preprint{AIP/123-QED}

\title{Controlling the spontaneous firing behavior of a neuron with astrocyte}
\author{Tugba Palabas}\affiliation{Department of Biomedical Engineering, Zonguldak Bulent Ecevit University, 67100, Zonguldak, Turkey}

\author{Andre Longtin}\affiliation{Department of Physics, University of Ottawa, Ottawa, Ontario, Canada}
	
\author{Dibakar Ghosh}\affiliation{Physics and Applied Mathematics Unit, Indian Statistical Institute, Kolkata 700108, India}
	
\author{Muhammet Uzuntarla*}\affiliation{Department of Bioengineering, Gebze Technical University, 41400, Kocaeli, Turkey}
\email{muzuntarla@yahoo.com}
\date{\today}

\begin{abstract}
		
		Mounting evidence in recent years suggests that astrocytes, a sub-type of glial cells, not only serve metabolic and structural support for neurons and synapses but also play critical roles in regulation  of proper functioning of the nervous system. In this work, we investigate the effect of astrocyte on the spontaneous firing activity of a neuron through a combined model which includes a neuron-astrocyte pair. First, we show that an astrocyte may provide a kind of multistability in neuron dynamics by inducing different firing modes such as random and bursty spiking. Then, we identify the underlying mechanism of this behavior and search for the astrocytic factors that may have regulatory roles in different firing regimes. More specifically, we explore how an astrocyte can participate in occurrence and control of spontaneous irregular spiking activity of a neuron in random spiking mode. Additionally, we systematically investigate the bursty firing regime dynamics of the neuron under the variation of biophysical facts related to the intracellular environment of the astrocyte. It is found that an astrocyte coupled to a neuron can provide a control mechanism for both spontaneous firing irregularity and burst firing statistics, i.e., burst regularity and size.
		
	\end{abstract}

\maketitle

\begin{quotation}
Astrocytes are the most numerous glial cells in the mature brain, and often surround neuronal somata and dendrites and provide fine enmeshment of synapses. Unlike neurons, these star-shaped cells do not elicit action potentials, yet they contribute to information processing via feedback to the cells by listening to the synaptic activity. We here investigate the their role on shaping the spontaneous firing behavior of a neuron. By using a neuron–astrocyte pair model, we identify distinct astrocyte-controlled neural activity patterns and explore their individual statistical characteristics under the variation of different biologically-plausible intrinsic astrocytic factors.

\end{quotation}	

\section{Introduction}
	
	Spontaneous electrical activity, observed in many neurons from distinct brain regions, has long been interested and is by definition the ongoing activity of the cells when they are not subjected to an input \cite{Dehaene2005, Redolf2021}. This non-stimulus-evoked activity can be observed in experiments by using different measuring tools at the scales of single neuron \cite{Kubska2021}, neural microcircuits \cite{Yuan2020} as well as whole brain \cite{Fox2007, Calim2021}. Depending on the brain region to which it belongs and many other biological factors, the spontaneous spiking of a neuron may appear with different temporal firing patterns that are broadly categorized as "repetitive'' \cite{Uddin2020}, "random" \cite{Paladini2003} and "bursty" \cite{Yang2018}. It is also possible to observe more complex patterns emerging with chaotic dynamics \cite{Yi2012, Dai2020}. Although it was initially considered as ''noise'' and various fundamental questions concerning the meaning and features of spontaneous neural activity are still a matter of debate, mounting evidence in recent years suggests that it may reflect molecular, cellular and network dynamics of the neurons and may play roles in specific neuronal computations \cite{Ozer2007, Uzuntarla2012, Frolov2020, Guiyang2021, Martini2021}.
	
	The spontaneous activity of a single neuron is more pronounced compared to that seen at the larger population scale. A widely observed typical feature of spontaneous spiking patterns is highly irregularity. In recent decades, there has been a great deal of effort in elucidating the functional implications of such irregularity both on the level of single cells and cell populations \cite{Ozer2006, Miura2007, Ozer2009, Payne2019}. Although the exact mechanism is not clear yet and the research is still in progress, the synaptic input correlations and the noisy environment have been suggested as the origin of the firing irregularity \cite{Brunel2003, Fellous2003, Burkitt2006, Mendonca2016}. In terms of its implications, many experimental and theoretical studies have reported that irregular spontaneous activity may reflect an underlying rich coding structure and it may be associated with several cognitive functions of the brain, such as working memory \cite{Hansel2013}, selective attention \cite {Ardid2010} and sensory coding \cite {Panzer2017}. Thus, the degree of irregularity in spontaneous firing patterns and its control may be critical to understand the underlying mechanisms of such cognitive functions or, more generally, information processing in the nervous system \cite{Kim2021}.
	
	To date, the control of irregular firing activity has been mostly linked to internal dynamics of neurons (i.e., excitability, ion channels, membrane currents) and background activity induced by inputs from neighboring neurons \cite{Dipoppa2013, Peterson2017}. However, the brain is not just composed of neurons but it also includes glial cells which are generally considered to provide metabolic and structural support for neurons and synapses \cite{Barber2019, Patel2019}. In recent years, many experimental studies have reported evidence that glial cells influence neural activity at both cellular and network level \cite{Jakel 2017, Allen2018}, but have not been studied in any detail before with respect to their role in control of temporal features of neural activity.	
	
	Astrocytes are the most numerous glial cells, are ubiquitous in the mature brain, and often surround neuronal somata and dendrites and provide fine enmeshment of synapses \cite{Sofroniew2010, Chung2015}. Findings from experimental studies suggest that they are involved in memory formation \cite{Zorec2015}, decision making \cite {Wang2017}, attention \cite{Guimaraes2018}, cognitive \cite{Brockett2018} and behavioral processes \cite{Hwang2021}, and modulate the behavior of neurons depending on their own physiological and environmental factors \cite{Halassa2010}. On the other hand, a growing body of evidence in recent years has demonstrated that the malfunctioning of astrocytes may contribute to various neurodegenerative diseases (i.e., Alzheimer, multiple sclerosis) \cite{Preman2021}. Unlike neurons, these star-shaped cells do not elicit action potentials, yet they contribute to information processing via feedback to the cells by listening to the synaptic activity. More precisely, the released neurotransmitter molecules at synapses during a spike transmission not only activate the target neuron but also they regulate $Ca^{2+}$ dynamics in astrocytes depending on the spike traffic at synapses. When the intensity of spike transmission at synapses is low, astrocytes respond with small $Ca^{2+}$ transients that cannot induce any significant input to the connected neurons. With the increase in the intensity of spike transmission at synapses, the intracellular $Ca^{2+}$ concentration exhibits oscillations, which constitutes the basis of the well-known $Ca^{2+}$ wave propagation \cite{Bazargani2016}, and triggers the release of several gliotransmitters (i.e., glutamate and adenosine) from astrocytes to the synaptic cleft. Pre-and postsynaptic neurons consider these gliotransmitter molecules as a feedback signal which regulates their internal dynamics \cite{ DePitta2009, yilmaz2019, calim2021}. 
	
	This signaling scheme in neuron–astrocyte communication suggests that astrocytes may be a significant factor in determining the level of neuronal spiking irregularity and may serve as a control mechanism \cite{Fellin2004, Wade2011}. Following this motivation, we here investigate the activity of a spontaneously firing neuron in a neuron–astrocyte pair, where the neuron is initially exhibiting spontaneous firing activity. More precisely, we identify distinct astrocyte-controlled neural activity patterns (eg. random, bursty) and explore their individual statistical characteristics under the variation of different biologically-plausible intrinsic astrocytic factors, i.e., the production rate of inositol trisphosphate, neuron-astrocyte coupling strength and the number of calcium channels.
	
\section{Models and Methods} 
	\label{sec:model}
	 We consider a coupled neuron-astrocyte pair with noise where the dynamics of the neuron is modeled with Fitzhugh-Nagumo ($FHN$) equations as follows \cite{FitzHugh1961, Yu2021}:
	
	\begin{equation}
		\varepsilon\frac{dV_{m}}{dt}=V_{m}-\frac{1}{3}V_{m}^{3}-\omega+\sqrt{2D}\xi(t),
	\end{equation}
	
	and
	
	\begin{equation}
		\frac{d\omega}{dt}=V_{m}+a_e+\lambda I_{astro},
	\end{equation}
		\noindent where $V_{m}$ denotes the electrical activity of the membrane and $\omega$ is the slow recovery variable that restores the resting state of the model. $\varepsilon$ is the time scaling parameter responsible for separation of fast ($V_{m}$) and slow ($\omega$) variables of the model, which is fixed to 0.005. $a_{e}$ is the bifurcation parameter that determines the excitability level of the neuron. Namely, the model neuron is in the excitable regime for $\mid a_{e} \mid$ $\geq 1$, with a single fixed point, while for $\mid a_{e} \mid < 1$, it exhibits oscillatory behavior. In the excitable regime, spontaneous activity of the neuron is caused by adding a Gaussian white noise source $\xi(t)$ with zero mean and intensity $D$ to the $FHN$ equations.
	
	The $I_{astro}$ term in Eq. (2) is a depolarizing astrocytic feedback current, that is scaled with the coupling constant $\lambda$, induced by the dynamical changes in the concentration of two main substances in the astrocyte: inositol trisphosphate ($IP3$) and cytosolic calcium ($Ca^{2+}$). Namely, an astrocyte responds to the neuronal spiking activity by binding released glutamate in the synaptic cleft to its metabotropic glutamate receptors ($mGluRs$). The activation of these receptors then triggers the production of $IP3$, which, consequently, causes the discharge of $Ca^{2+}$ internal stores mediated by $IP3$ receptor channels ($IP3Rs$). Such signaling scheme results in alteration of $Ca^{2+}$ concentration in the astrocyte cytosol. To model these processes, we use the approximation proposed by Nadkarni and Jung \cite{nadkarni2003} for the dynamics of $IP3$ production in the astrocyte as follows:
	
	\begin{equation}
		\frac{d[IP3]}{dt}=\frac{1}{\tau_{IP3}}([IP3]^{*}-[IP3])+r_{IP3} \frac{1}{1+e^{(\theta_{s}-V_{m})/\sigma_{s}}},
	\end{equation}
	
	\noindent where the first term on the right hand side refers to decay in intracellular $IP3$ concentration whereas the second one, defined by a sigmoid function with steepness $\sigma_{s}$, refers to its production which is triggered when $V_{m}$ exceeds the threshold voltage $\theta_{s}$. $IP3^{*}$=160nM and $\tau_{IP3}=0.00014 ms^{-1}$ are the experimentally determined values for the equilibrium concentration of $IP3$ and the decomposition time constant, respectively. The parameter $r_{IP3}$ denotes the astrocyte $IP3$ production rate in response to a single spike emitted by the neuron \cite{amiri2011}.
	
	The intracellular $Ca^{2+}$ concentration is modeled based on the well-known Li-Rinzel ($LR$) equations which describe the $Ca^{2+}$ dynamics with three distinct fluxes: $J_{channel}$ is $Ca^{2+}$ flux from $ER$ (Endoplasmic Reticulum) to the cytosol through $IP3R$ channels, $J_{pump}$ is the $Ca^{2+}$ flux from cytosol into $ER$ via $ATP$-dependent pumps and $J_{leak}$ is the leakage flux from the $ER$ to intracellular space due to difference in $Ca^{2+}$ concentrations between ER and cytosol. The model equations for the $Ca^{2+}$  dynamics are as follows \cite{lirinzel1994, manninen2018}:
	
		\begin{equation}
		\frac{d[Ca^{2+}]}{dt}=-J_{channel}-J_{pump}-J_{leak},
	\end{equation}
	
\noindent with fluxes defined as
		
	\begin{equation}
		J_{channel}=c_{1}v_{1}m_{\infty}^{3}n_{\infty}^{3}q^{3}([Ca^{2+}]-[Ca^{2+}]_{ER}),
	\end{equation}
	\begin{equation}
		J_{pump}=\frac {v_{3}([Ca^{2+}]^{2})}{k_{3}^{2}([Ca^{2+}]^{2})},
	\end{equation}
	\begin{equation}
		J_{leak}=c_{1}v_{2}([Ca^{2+}]-[Ca^{2+}]_{ER}),
	\end{equation}
	
	where
	\begin{equation}
		m_{\infty}=\frac {[IP3]} {[IP3]+d_{1}},            \qquad            n_{\infty}=\frac {[Ca^{2+}]} {[Ca^{2+}]+d_{5}},  
	\end{equation}
	
	and
	
	\begin{equation}
		[Ca^{2+}]_{ER}=\frac {c_{0}-[Ca^{2+}]} {c_{1}}.              
	\end{equation}
	
	Here, $q$ refers to the fraction of activated $IP3Rs$ and satisfies the equation
		\begin{equation}
		\frac{dq}{dt}=\alpha_{q}(1-q)-\beta_{q}q+\xi_{q}(t),
	\end{equation}
where $\alpha_{q}$ and $\beta_{q}$ refer the activation and inactivation rates of these receptor channels (see \cite{nadkarni2003} for the equations). In our study,  stochastic dynamics for $IP3Rs$ are also taken into account, and modeled with the $Langevin$ approach where random kinetics of the $IP3Rs$ are incorporated into model via an independent zero-mean Gaussian white  noise source $\xi_{q}(t)$ whose autocorrelation function is defined as \cite {tang2016}
	
	\begin{equation}
		\langle \xi_{q}(t)\xi_{q}(t^{'})\rangle=\frac {\alpha_{q}(1-q)\beta_{q}q}{Nt}\delta(t-t^{'}),
	\end{equation}
	
	\noindent where  $Nt$ corresponds to the number of $IP3Rs$ which determines the channel noise strength. Note that $Nt$ and effective channel noise level are inversely related. Unless stated otherwise, the channel noise is ignored throughout our analysis in this work to understand the fundamental principles of communication between neuron and astrocyte. The values of  astrocyte parameters are listed in Table I \cite{Volman2007, Postnow2009}.\\
	
	Finally, the astrocytic current equation is obtained from Nadkarni and Jung \cite{nadkarni2003} as a function of intracellular $Ca^{2+}$ concentration of the astrocyte according to experimental data defined by
	\begin{equation}
		I_{astro}=2.11ln(y)\Theta(lny), \qquad  y=[Ca^{2+}]-196.69
	\end{equation}
	
	\noindent where $I_{astro}$ is  the slow inward current which acts as a depolarizing input to the neuron. $y$ is the amount of $Ca^{2+}$ above the threshold level and $\Theta$ is a Heaviside function: $H(x)=1\mbox{ for } x>0, \mbox{ and } H(x)=0 \mbox{ for } x<0$. For statistical accuracy, each data point in the following results is obtained by averaging 50 independent realizations of the model equations for any given set of the neuron and astrocyte parameters (noise level, coupling strength, $IP3$ production rate).\\
	
	\begin{table}[t]
		\caption{Astrocyte standard parameters }
		\label{parameter}
		\begin{tabular}{|l|l|l|}
			\hline
			\textbf Parameter & \textbf Value & \textbf Description \\ \hline
			$c_{1}$ & $0.185$ & Ratio of ER volume to cytosol volume\\ \hline
			$v_{1}$ & $6 sec^{-1}$ & Maximum $ Ca^{2+}$ channel flux\\ \hline
			$v_{2}$ & $0.11 sec^{-1}$ &$ Ca^{2+}$ leak flux constant\\ \hline
			$v_{3}$ & $0.9 \mu M sec^{-1}$ & Maximum  $Ca^{2+}$ uptake\\ \hline
			$k_{3}$ & $0.1 \mu M$ & SERCA activation constant\\ \hline
			$d_{1}$ & $0.13 \mu M$ & $IP3$ dissociation constant\\ \hline
			$d_{5}$ & $0.08234 \mu M$ & $Ca^{2+}$  activation dissociation constant\\ \hline
			$c_{0}$ & $2.0 \mu M$ & Cytosolic free $Ca^{2+}$ concentration \\ \hline
		\end{tabular}
	\end{table}
	
\section{Results}
	
	In what follows, we will systematically investigate the effects of astrocytes on the firing dynamics of a single neuron by considering the proposed neuron-astrocyte coupled model. To do so, we begin by exploring how an astrocyte changes the spontaneous spiking behavior of the proposed neuron model by changing the coupling strength $\lambda$. In Fig.1, we present representative voltage traces of membrane as a function of time for different $\lambda$ values of the astrocyte-neuron pair. It is seen in the top panel that the isolated neuron ($\lambda=0$) exhibits noise-induced firing activity. When the effect of the astrocyte is considered, i.e., $\lambda>0$, dramatic variation in firing profile and new firing regimes start to emerge with the increased values of $\lambda$ as illustrated in the following panels of Fig. 1. For instance, when the coupling is weak, we observe that neuronal firing activity decreases during some specific periods of time. Further increase in $\lambda$ results in a significant reduction in the number of spikes (see panels c and d) and, finally, the activity completely disappears during these periods (see panels e, f and g) when $\lambda$ exceeds a certain level. These observations suggest that an astrocyte provides a kind of multistability into the neural dynamics by switching the firing behavior from one mode to the another, i.e., from random spiking to bursty firings or from silence to random spiking.

	\begin{figure}[ht!]
		\includegraphics[trim=0.0cm 0cm 0.0cm 0cm, clip=true, scale=0.5]{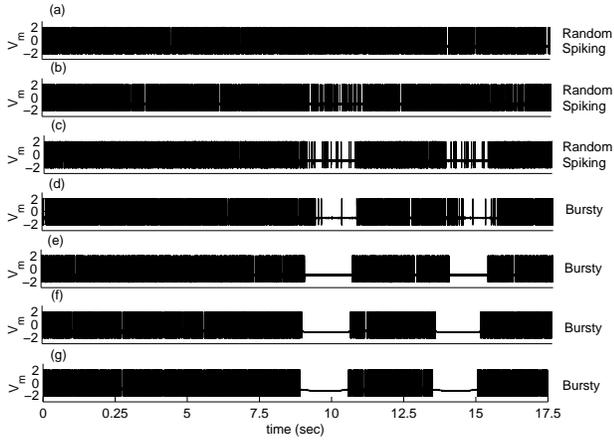}
		\caption{ Firing behavior of the model neuron in the excitable regime with and without astrocyte interaction. All panels show time evolution of the neuron's membrane potential for a fixed parameter set as $a_{e}=1.01$, $D=0.04$ and $ r_{IP3}=1.5$. (a) $\lambda=$0 (without astrocyte interaction), (b) $\lambda=0.00025$, (c) $\lambda=0.00075$, (d) $\lambda=0.001$, (e) $\lambda=0.0025$, (f) $\lambda=0.0075$, (g) $\lambda=0.025$. To distinguish the random and bursty spiking modes, we apply a rate-threshold method during quiescient periods of bursting activity. When the average number of background activity induced firings during these periods is higher than a threshold value (50 spikes/s), the firing mode of the neuron is determined as random spiking. Otherwise, we consider the firing mode as bursty.}
	\end{figure}
	
	Next, we explore the effect of noise intensity $D$ for different values of coupling strength $\lambda$ and excitability level $a_e$. By changing these parameters, we observe three different neuronal phenomena, namely silence, bursty and random spike states. To classify our observations in Fig. 1, we perform a phase plane analysis in our neuron-astrocyte coupled system. Fig. 2 presents the activity mode of the neuron on ($D$, $\lambda$) plane for three different excitability levels ($a_{e}$). It is seen that there exists three different activity mode regardless of the excitability level of the neuron: silence, random spiking (RS) and bursty behavior. Note that these activity profiles at different excitability states emerge due to the inputs from both astrocyte and noise. It is seen that if the $D$ is not high enough than a certain value, the neuron is silent for all $a_{e}$ values. The critical level of $D$ shifts to the right (large noise) as $a_{e}$ increases. This is mainly due to the fact that the closer values of $a_{e}$ to the bifurcation point ($a_{e}=1.0$) in the FHN model neuron provides more excitability where the occurrence of a spike requires less noise. On the other hand, when $D$ is sufficiently large, the neuron exhibits RS activity as well as bursty behavior. But the latter one occurs at relatively large values of $D$ and $\lambda$ and the region of this firing regime shrinks as $a_{e}$ increases. These observations on the phase plane suggest that it is possible to switch firing mode of a neuron by fine tuning $\lambda$ and $D$ without any change in its internal dynamics.

	\begin{figure}[ht!]
		\centerline{
		\includegraphics[trim=0.0cm 0cm 0.0cm 0cm, clip=true, scale=.4]{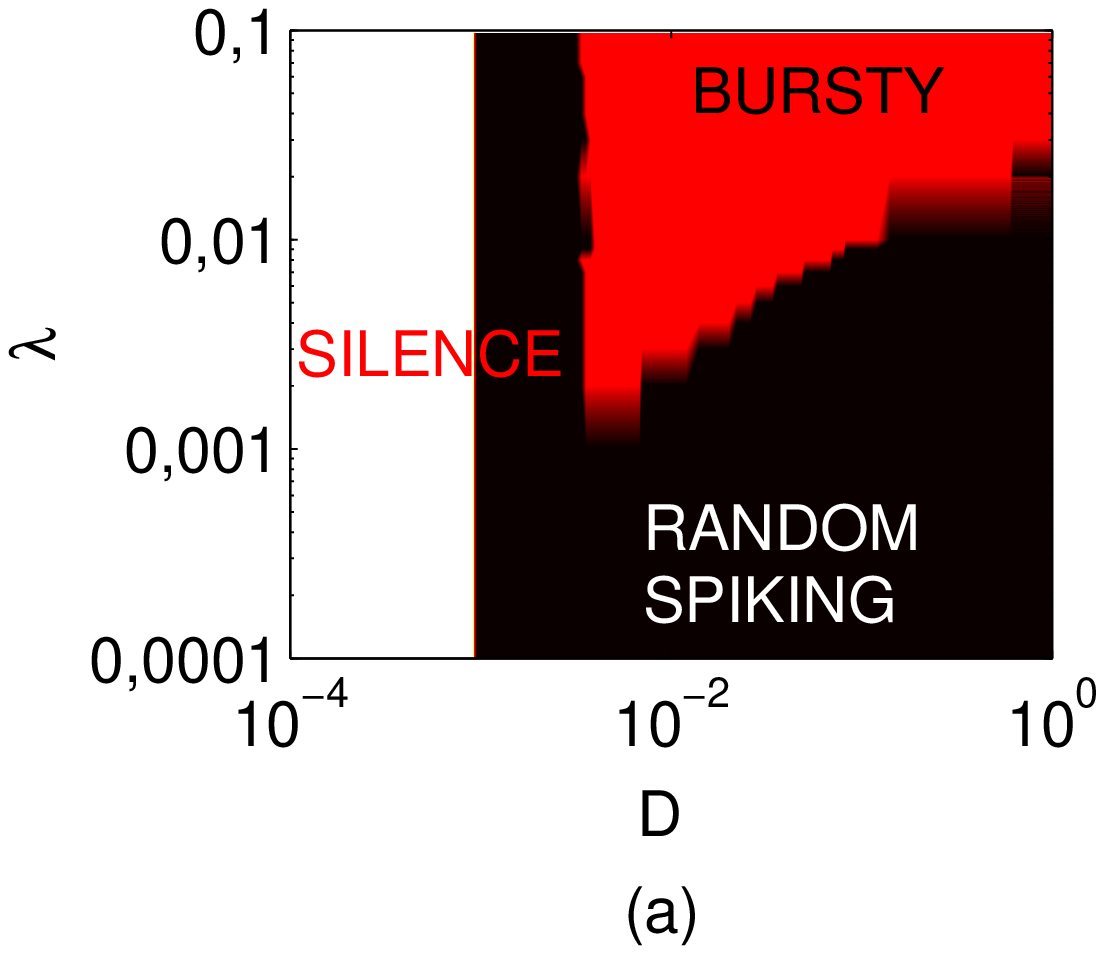}
		\includegraphics[trim=0.0cm 0cm 0.0cm 0cm, clip=true, scale=.4]{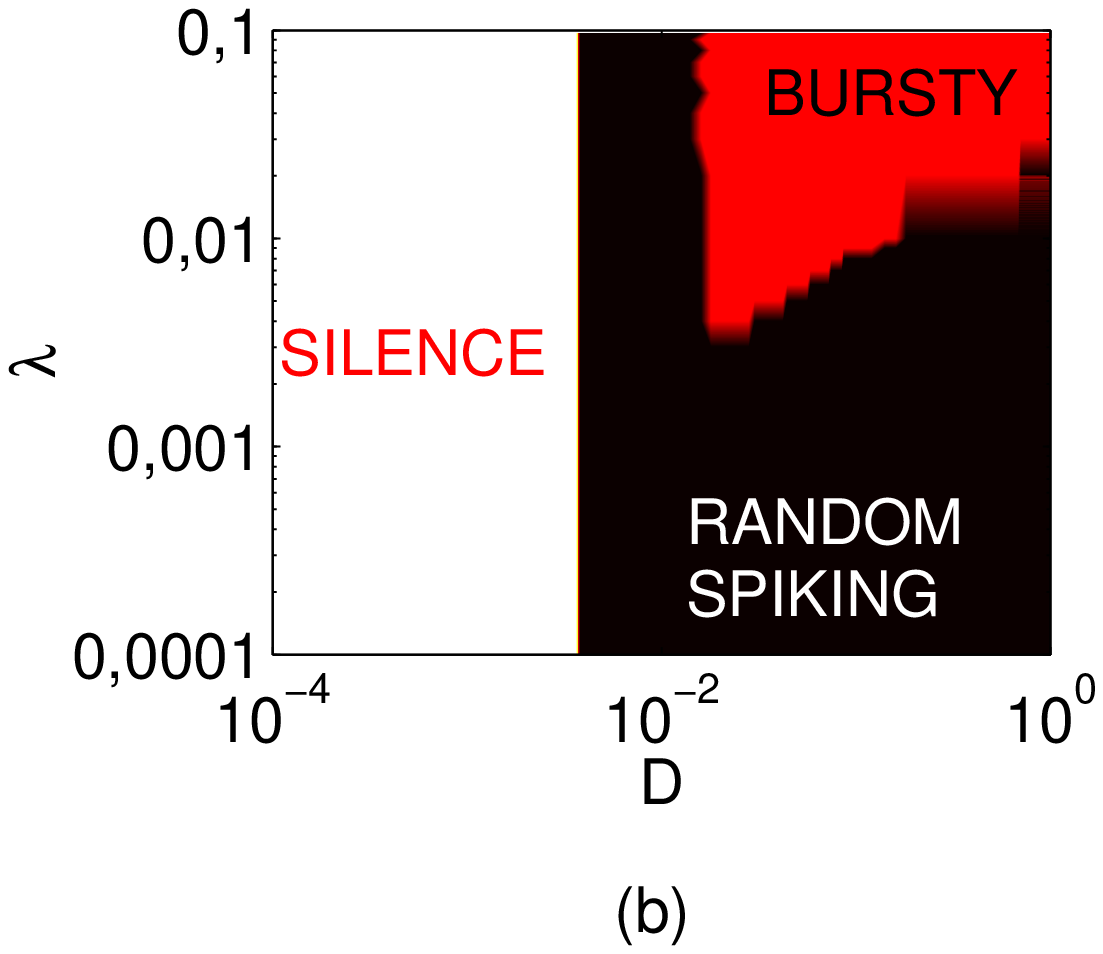}}
		\includegraphics[trim=0.0cm 0cm 0.0cm 0cm, clip=true, scale=.4]{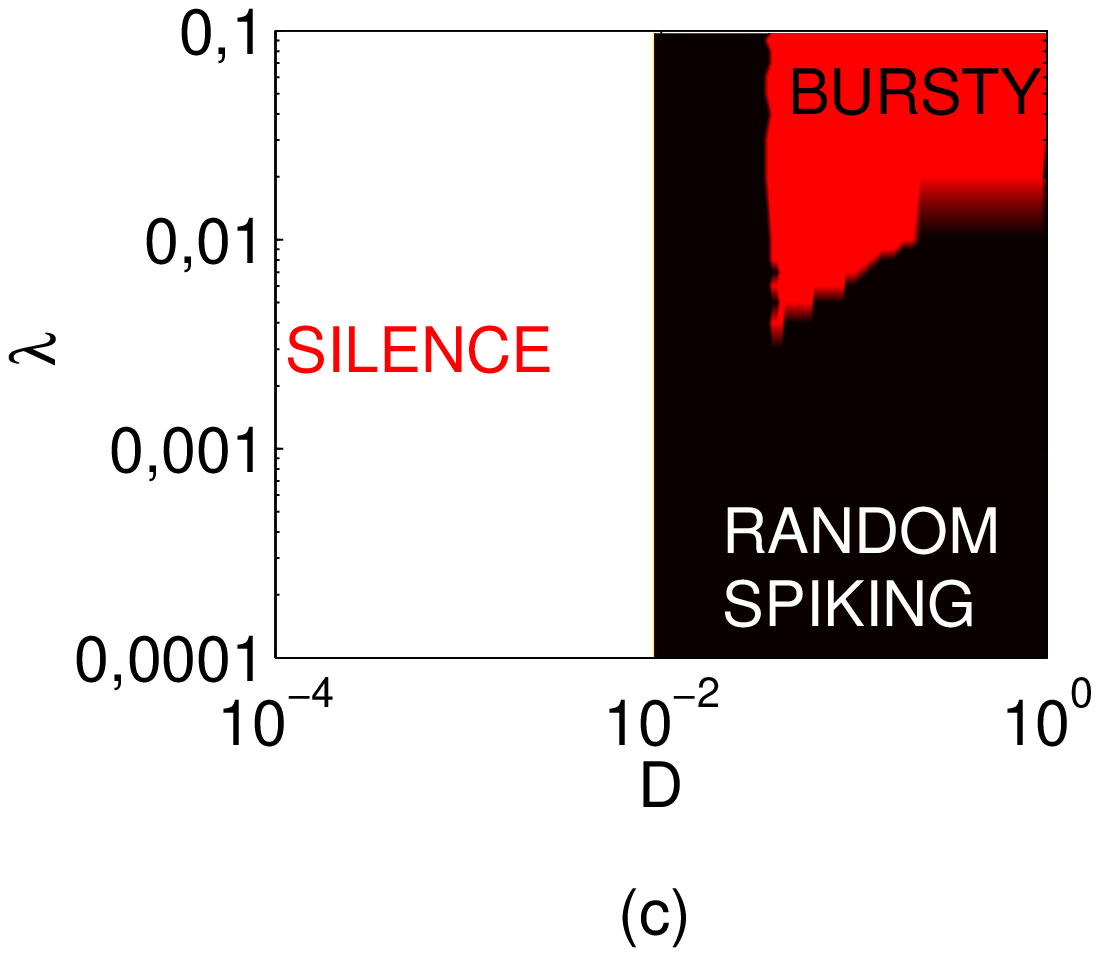}
		\caption{ Each panel shows phase diagrams spanned by parameters $D$ and $\lambda$ for different excitability level of the neuron: (a) $a_{e}=1.01$, (b) $a_{e}=1.03$, (c) $a_{e}=1.05$. Regions of different firing behavior are given by the following color codes, Black: Random spiking regime, Red (gray): Bursty Regime, White: Silence. The rate of $IP3$ production is set to $r_{ip3}=1.5$. In these parameter regions, the region of silence state is enlarged and the bursty state observed for higher noise intensity and excitability level $a_e$.}
	\end{figure}
	
Following our initial motivation, we now continue to analyze the effect of astrocyte on the firing behavior of the neuron by considering the facts observed in the above phase-plane analysis. Since the prominent features of random spiking and bursty modes differ from each other, separate analyses for these two firing regimes are necessary. More precisely, we investigate the impact of astrocyte in determining the neuron's spiking (ir)regularity which is believed to be important for information processing capacity of neurons. On the other hand, for the case of bursty mode, we study the role of neuron-astrocyte communication on the frequency and size of the bursts.
	
	\subsection {Spontaneous spiking regularity with astrocyte}
	
	Both \textit{in vivo} and \textit{in vitro} recordings indicate that spontaneous neural spiking activity in most brain regions is highly irregular \cite{Shadlen1998, Zierenberg2018, Uva2021}. Over the past recent decades, understanding the nature, cause and functionality of that irregularity has been attracted a great deal of attention and many hypotheses have been proposed regarding these issues. For instance, it has been suggested that the irregular timing of spikes could convey information, providing a broad information bandwidth on the neural spike train \cite{Dettner2016, Ventura2019}. On the other hand, the irregularity may be the reflection of noise which limits the information processing capacity of cells \cite{ Pisarchik2019, Tischbirek2019}. So far, in terms of nature and cause, although various physiological mechanisms including inherent neural dynamics and environmental factors (e.g. stochastic dynamics of ion channels, balance between excitation and inhibition) have been proposed as the source and the controller of spiking irregularity, the underlying mechanism for this phenomenon is still not known exactly \cite{He2019, Ushakov2021, Xu2021}. Thus, we are interested in what would be the role of astrocytes in determining such irregular firing patterns. 

	\begin{figure}[htp]
		\centerline{
		\includegraphics[trim=0.0cm 0cm 0.0cm 0cm, clip=true, scale=.4]{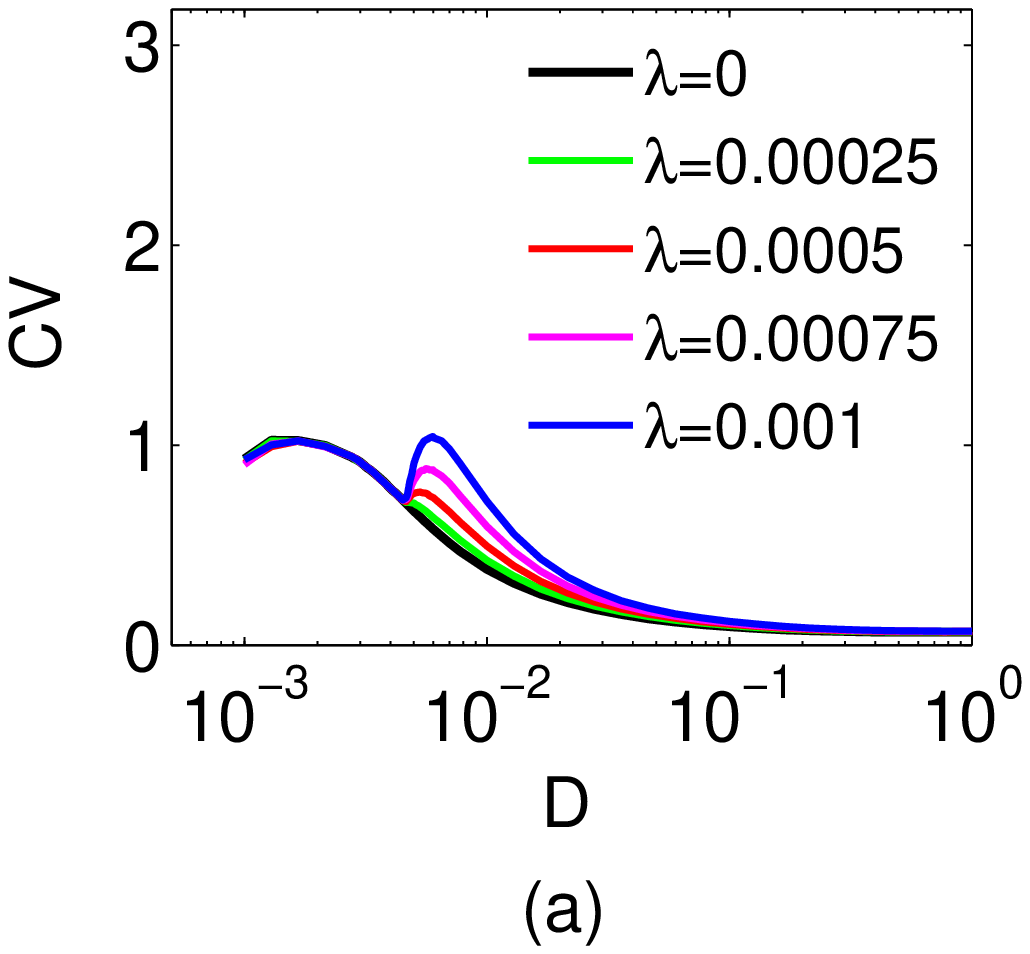}
		\includegraphics[trim=0.0cm 0cm 0.0cm 0cm, clip=true, scale=.4]{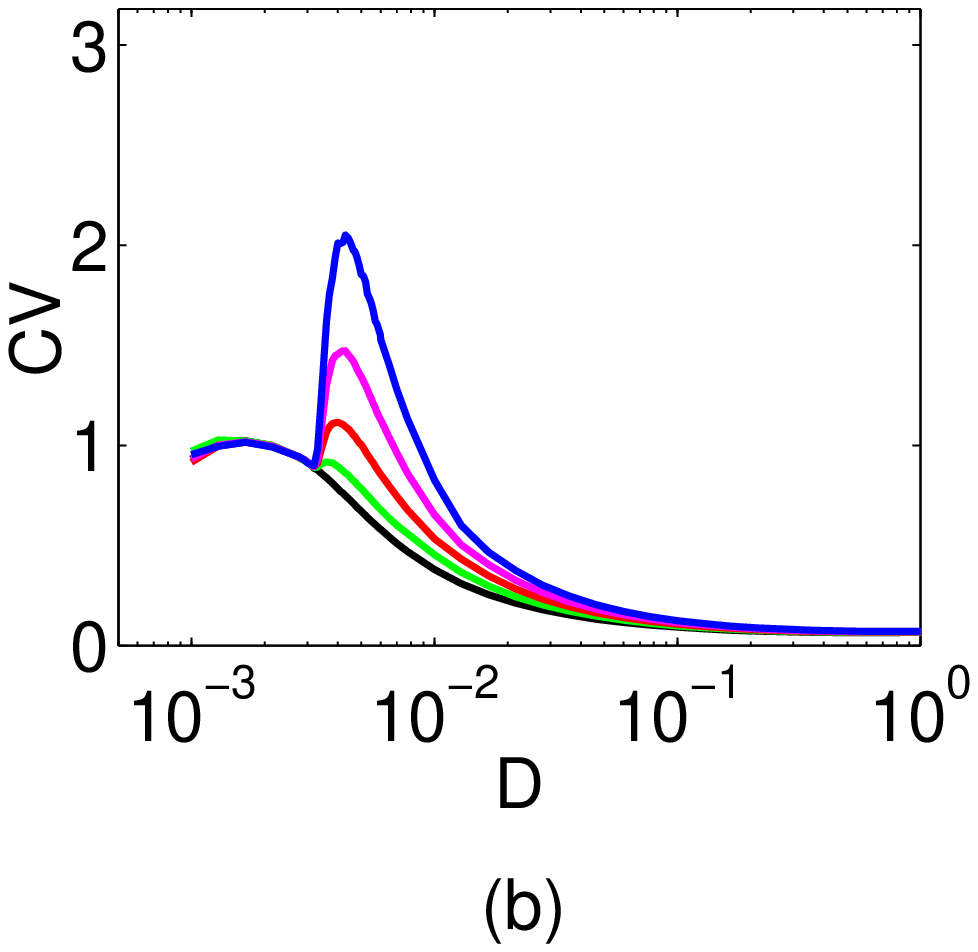}}
		\includegraphics[trim=0.0cm 0cm 0.0cm 0cm, clip=true, scale=.4]{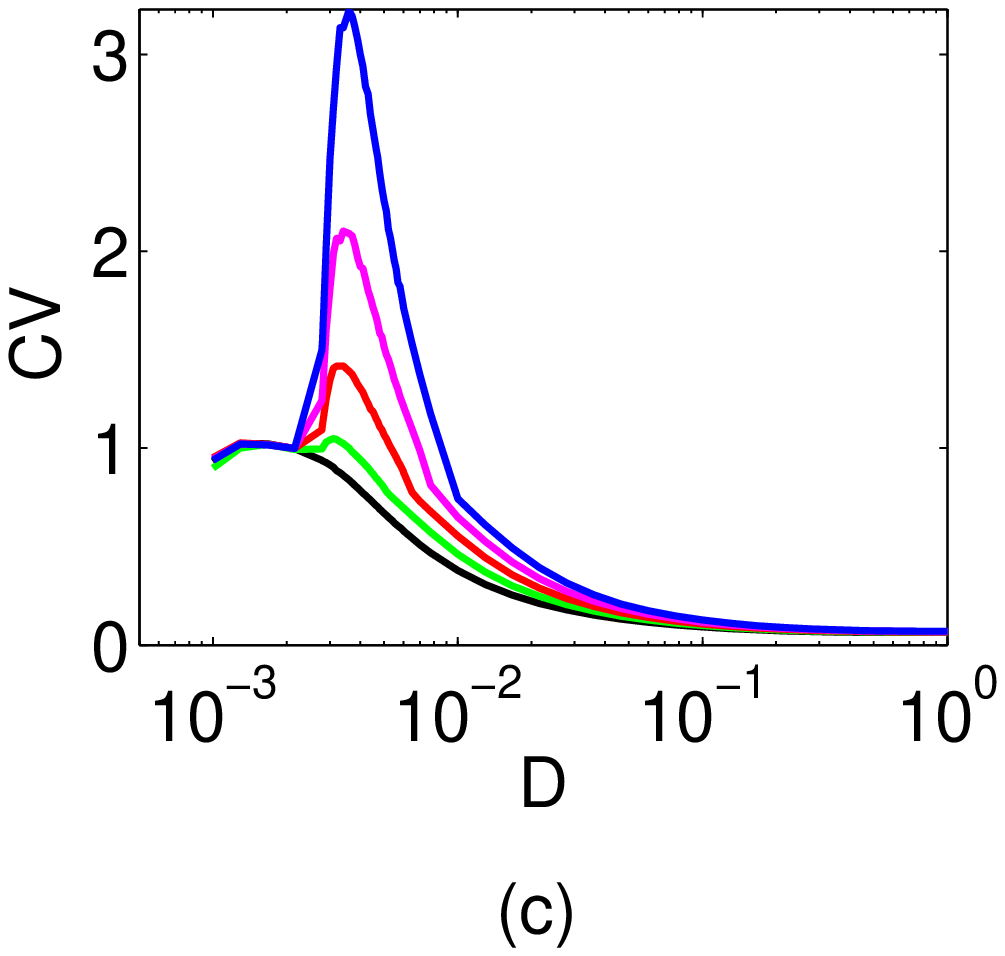}
		\caption{The dependence of the CV, a measure of spontaneous spiking irregularity, on $D$ and $\lambda$ for different values of $r_{ip3}$: (a) $r_{ip3}=0.8$, (b) $r_{ip3}=1.5$, (c) $r_{ip3}=2.5$. The presented results are obtained for a fixed level of excitability of the neuron, which is set to $a_{e}=1.01$.}
	\end{figure}
	
	To do so, we simulate our coupled model systematically and search the astrocytic factors that may have effects on neural activity of the neuron through the $CV$ (Coefficient Variation) of spike trains. $CV$ is a widely used measure of spiking irregularity which is obtained by the division of the standard deviation to mean of the interspike intervals ($ISIs$) and defined by \cite{Schmid2001, Ozer2009}:

	\begin{equation}
		CV=(\frac{\sqrt{\sum(ISI_{i}-\mu)^{2}/N-1}}{\sum ISI_{i}/N}; \qquad i=1,2,...,N
	\end{equation}

 It should be noted that the larger values of the $CV$ imply more irregular spike trains. First, we investigate how $CV$ changes with the coupling strength $\lambda$ as a function of $r_{ip3}$ and $D$ for a fixed excitability level of the neuron $a_{e}= 1.01$. Obtained results are presented in Fig. 3. It is seen that $CV$ first exhibits a small increase for all $\lambda$ at lower values of $D$ where the astrocyte does not provide any significant feedback to the neuron due to the lack of required number of arriving spikes for the calcium accumulation. More precisely, such noise levels are not enough to turn on the astrocyte, and this small increase in $CV$ is not a consequence of neuron-astrocyte interaction. It appears due to the effect of noise on internal dynamics of $FHN$ neuron. On the other hand, for moderate noise levels, a second resonance peak emerges with the activation of feedback from astrocyte to the neuron where the $CV$ increases as $\lambda$ get larger values. Finally, for very large values of $D$, the noise suppresses the inputs from astrocyte and the neuron fires under the influence of noise, which results in an overlap of $CV$ curves for all neuron-astrocyte coupling strengths, following the $CV$ trend of an isolated neuron. Fig. 3 also features the effect of $IP3$ production rate on $CV$ behavior. It is seen that the above mentioned coupling-induced resonance peaks become more pronounced at all $\lambda$ levels with the increased values of $r_{ip3}$ (see maximum $CV$ values for a given $\lambda$ in all panels). Moreover, it is also obvious that resonance curves shift to the left for higher $r_{ip3}$, indicating that production rate provides less noise requirement for higher resonance peaks in $CV$.

	The above results clearly demonstrate that inherent dynamics of an astrocyte can induce more irregular firing patterns of spontaneous spiking activity for a particular range of noise intensity as well as provide a control mechanism for the level of such irregularity. To verify the validity of this conclusion and to investigate the dependency of astrocytic control on internal dynamics of the neuron, we perform similar analyses for different values of $a_{e}$ since the level of neuronal excitability is critical in determining the required spike inputs to activate the astrocyte. Obtained results are shown in Fig. 4, demonstrating the $CV$ versus $D$ and $a_{e}$ for four different values of $r_{ip3}$. In the case of isolated neuron (see Fig. 4a for $r_{ip3}=0$), since the astrocyte does not provide any feedback to the cell, variation of $CV$ curves are determined only by the noise and $a_{e}$. It is seen that similar $CV$ trends occur for different excitability levels, which only shift to the right for increasing values of $a_{e}$. This is due to the fact that a less excitable neuron (i.e., large $a_{e}$ ) needs more noise to fire an action potential. On the other hand, when $r_{ip3}>0$ (see Fig. 4b-d), we observe the similar effect as in our previous analysis in Fig. 3 where the required noise intensity for resonance peaks decreases as $r_{ip3}$ increases at all excitability levels. One can easily follow this shift by comparing $CV$ curves for a given excitability level in panels of Fig. 4. It is also obvious that there exists a threshold level of  $r_{ip3}$ for different excitability levels of the neuron to effectively control the $CV$. For instance, when $r_{ip3}=0.8$ (see Fig. 4b), the above mentioned astrocyte-induced second resonance peak of $CV$ starts to emerge with decreasing amplitudes as $a_{e}$ increases. With the increase in $r_{ip3}$, the second resonance peaks for all considered $a_{e}$ values become clearly pronounced. Fig. 5 provides a better illustration for the influence of such interaction between neuron and astrocyte parameters where we only plot maximum $CV$ ($CV_{max}$) values computed in Fig. 4 as a function of $a_{e}$ and $r_{ip3}$. It is seen that $CV_{max}$ is independent from $a_{e}$ when $r_{ip3}$ is less than a threshold value which is around $0.8<r_{ip3}<1.5$. These findings indicate that a less excitable neuron needs astrocyte partners having large $IP3$ production rate to exhibit more irregular firing activity.

	\begin{figure}[htp]
		\centerline{
		\includegraphics[trim=0.0cm 0cm 0.0cm 0cm, clip=true, scale=.35]{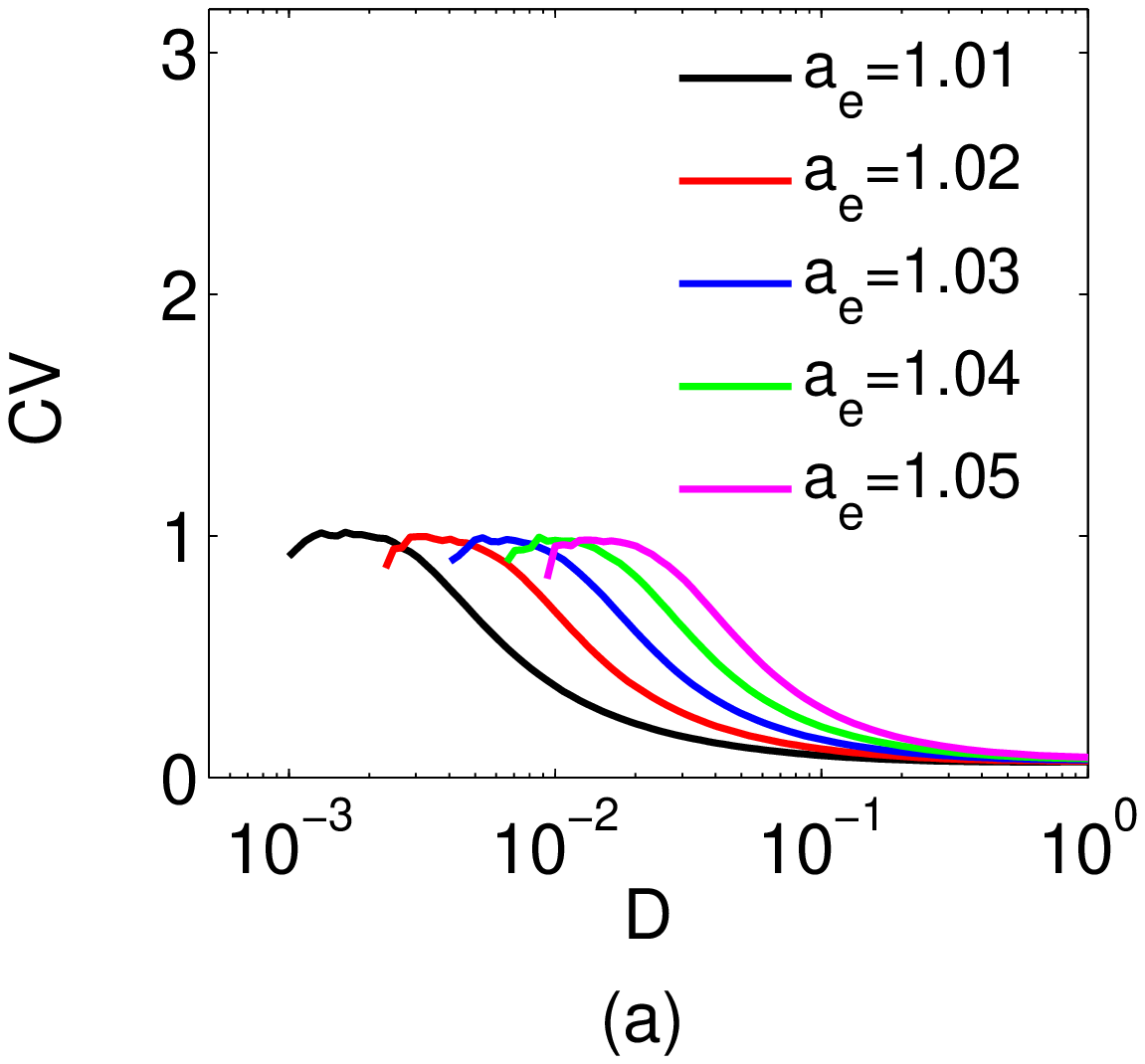}
		\includegraphics[trim=0.0cm 0cm 0.0cm 0cm, clip=true, scale=.35]{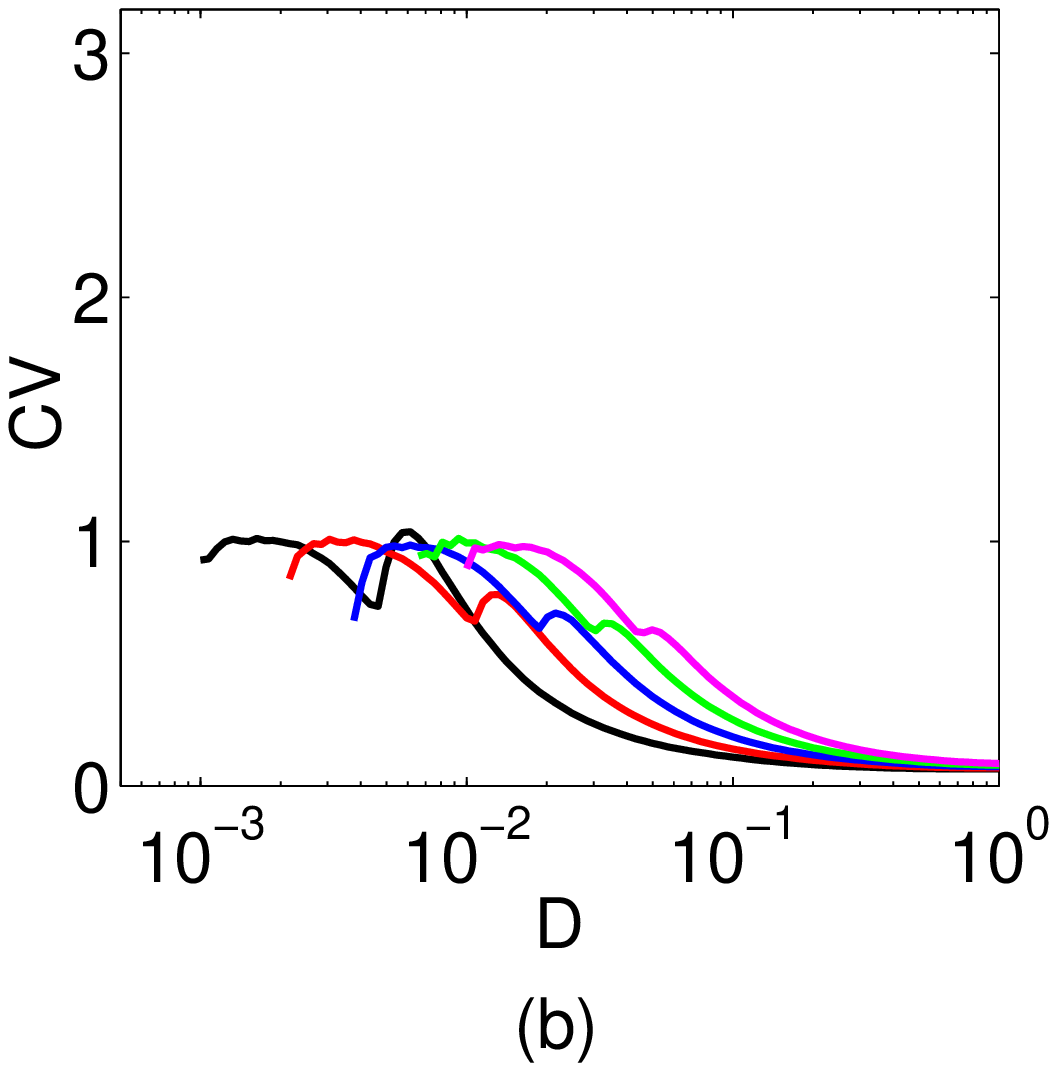}}
		\centerline{
			\includegraphics[trim=0.0cm 0cm 0.0cm 0cm, clip=true, scale=.35]{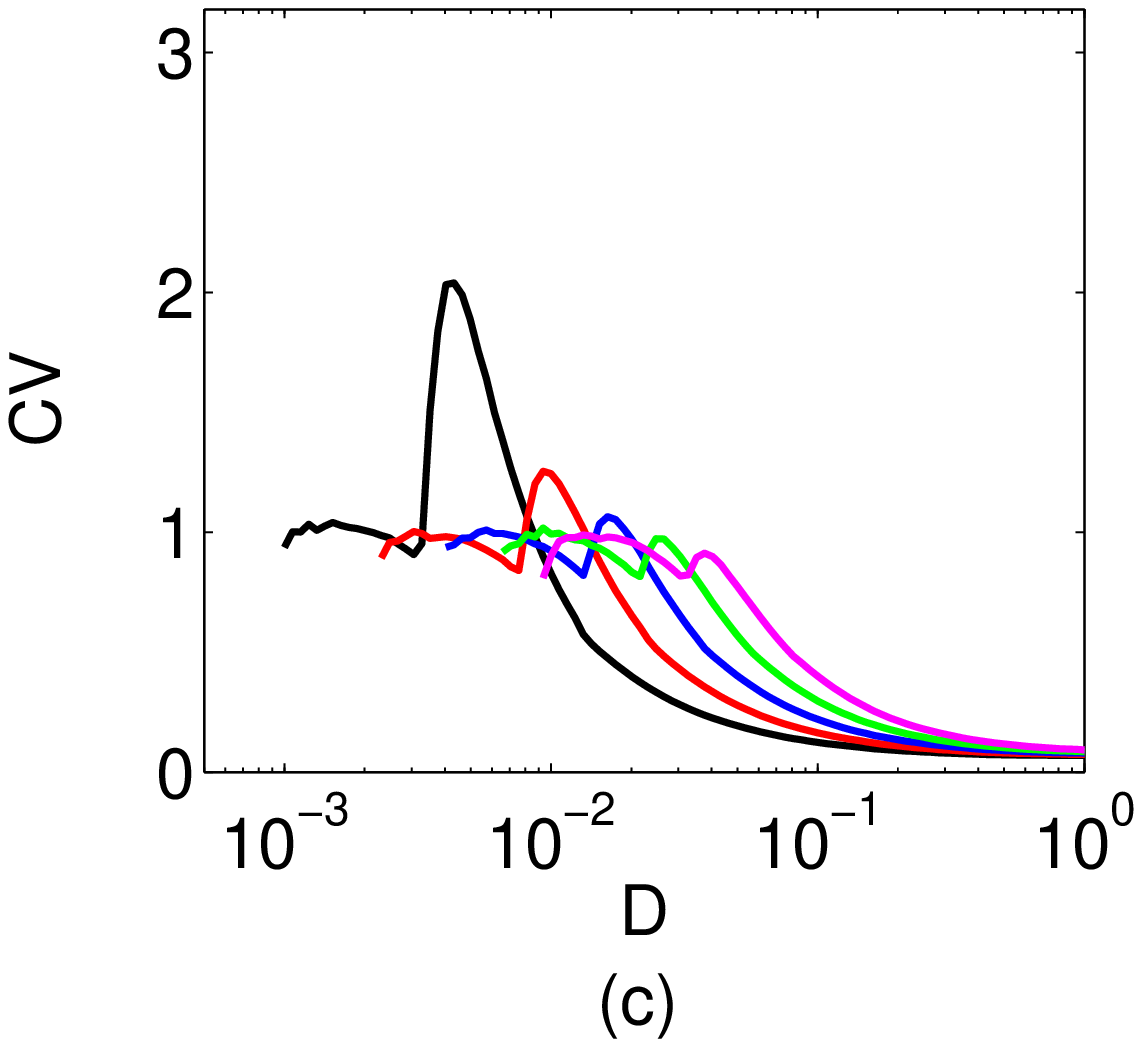}
		\includegraphics[trim=0.0cm 0cm 0.0cm 0cm, clip=true, scale=.35]{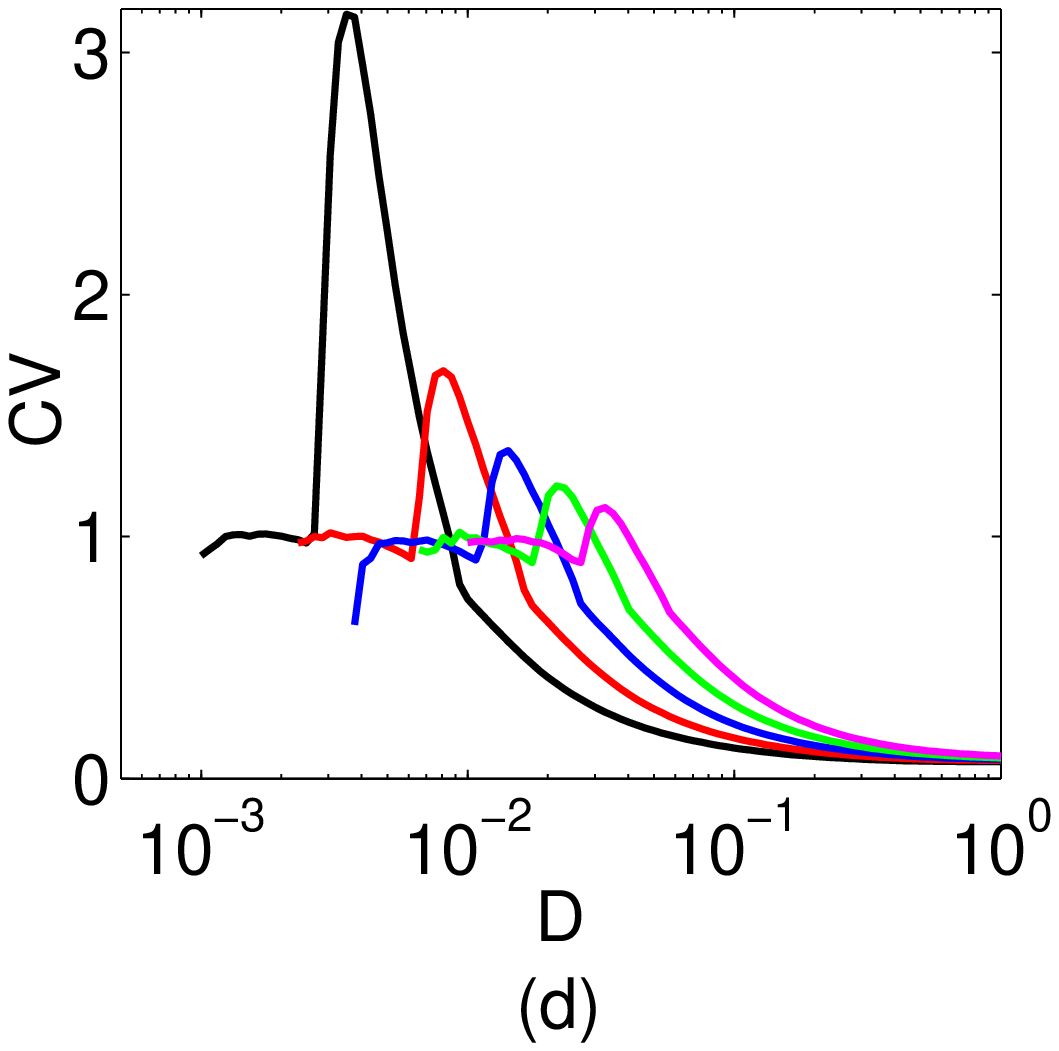}}
		\caption{ The dependence of the spontaneous spiking irregularity as a function of $D$ and $a_{e}$ for different values of $r_{ip3}$, (a) $r_{ip3}=0$, (b) $r_{ip3}=0.8$, (c) $r_{ip3}=1.5$, (d) $r_{ip3}=2.5$. The presented results are obtained for a constant coupling strength which is set to $\lambda=0.001$. }
	\end{figure}

	\begin{figure}[htp]
		\includegraphics[trim=0.0cm 0cm 0.0cm 0cm, clip=true, scale=.5]{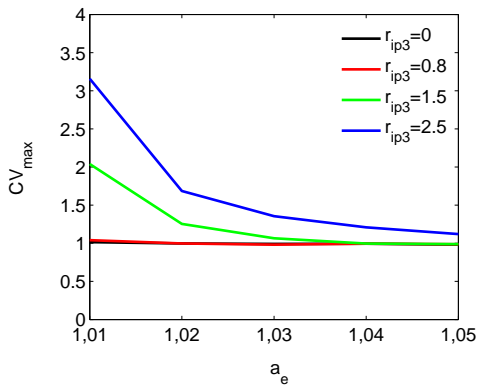}
		\caption{Variation of the maximum level of spontaneous spiking irregularity as a function of $a_{e}$ for different values of $r_{ip3}$. The presented results are obtained for a fixed astrocyte-neuron coupling strength $\lambda=0.001$.}
	\end{figure}
	
\begin{figure*}[ht!]
		\includegraphics[trim=0.0cm 0cm 0.0cm 0cm, clip=true, scale=.46]{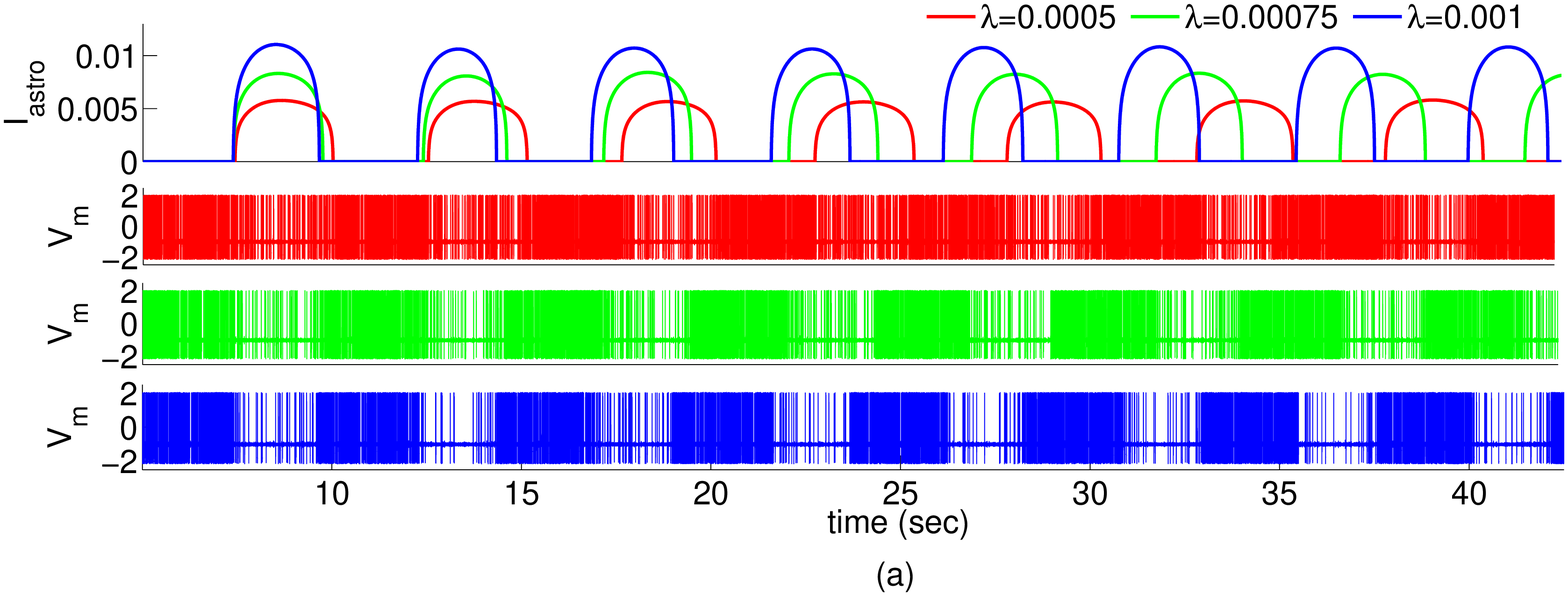}
		\includegraphics[trim=0.0cm 0cm 0.0cm 0cm, clip=true, scale=.46]{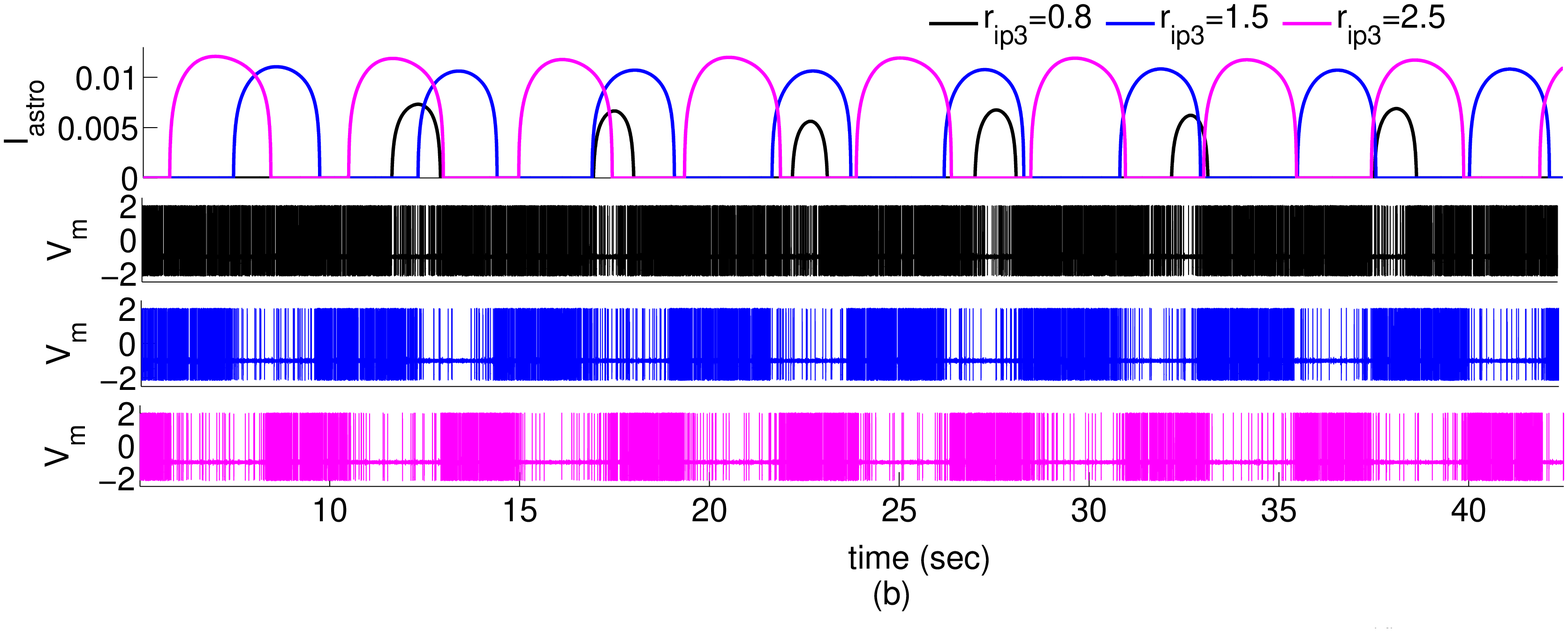}
		\caption{Time variation of the membrane potential of the neuron and the feedback current $I_{astro}$ as functions of intrinsic control parameters of the astrocyte. (a) The top panel shows $I_{astro}$ versus time with three different values of $\lambda$ for a fixed $r_{ip3}=1.5$ and the following panels present the corresponding voltage traces of the neuron for the color coded $I_{astro}$ feedback currents. (b) The same analysis as in (a) with the variation of $r_{ip3}$ for a fixed value of $\lambda=0.001$. The other parameters of the model are $a_{e}=1.01$ and $D=0.005$.}
	\end{figure*}

	To explain the mechanism underlying neuron-astrocyte cross-talk and improve our understanding on the analysis presented above, we demonstrate the traces of the neuron's membrane voltage and the astrocyte feedback current $I_{astro}$ for different values of $\lambda$ and $r_{ip3}$ (see Fig.6). Note that the color-coded $I_{astro}$ traces in the top panels of Fig. 6 (a) and (b) correspond to the same color-coded membrane potential traces shown in the following panels of each figure. It is seen that the lifetime and the amplitude of the $I_{astro}$ significantly change depending on the variations in $\lambda$ and $r_{ip3}$. Namely, the increase in both parameters results in amplitude enhancement as expected. However,  as we increase $\lambda$ and $r_{ip3}$, the lifetime period decreases and increases, respectively. By observing the corresponding voltage traces in the following panels, one can see that there is a relevant correlation between the firing activity and $I_{astro}$ where the number of spontaneous firings decrease when $I_{astro}$ is strong enough. The activity reduction periods become more pronounced for increased values of both coupling strength $\lambda$ and $r_{ip3}$. It should be noted that a further increase of these astrocytic parameters induces complete cessation of firing activity during the periods of time when the astrocyte is active (data not shown). This bursty behavior will be discussed in the next section. The firing activity reductions at random periods of time arise from the fact that the feedback current $I_{astro}$ decreases the excitability of the neuron. Therefore, the more frequent and long-lasting disruptions of firing activity by $I_{astro}$ induce more heterogeneous interspike interval distributions resulting in increased levels of spontaneous spiking irregularity. The results concerning the regulatory role of astrocyte on spontaneous spiking statistics presented in Figs. (1-5) can be interpreted with this understanding.
		\begin{figure}[ht!]
		\includegraphics[trim=0.0cm 0cm 0.0cm 0cm, clip=true, scale=.5]{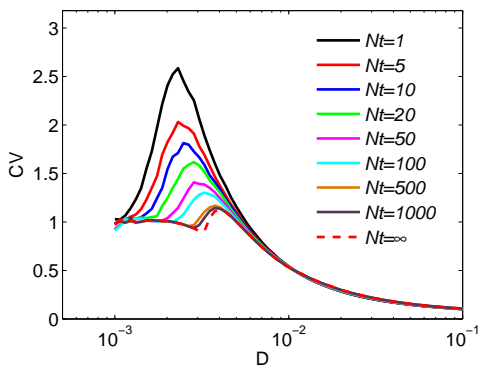}
		\caption{ The dependence of the spontaneous spiking irregularity on $D$ for different values of $Nt$. The presented results are obtained for a fixed parameter sets, which are $a_{e}=1.01$, $\lambda=1.01$ and $r_{ip3}=1.5$.}
	\end{figure}

\begin{figure*}[ht!]
		\includegraphics[trim=0.0cm 0cm 0.0cm 0cm, clip=true, scale=.48]{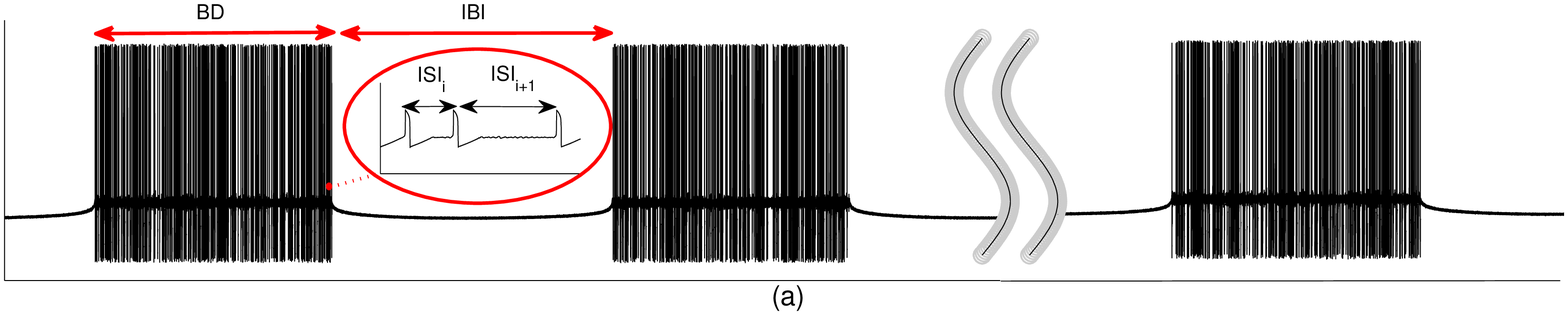}
		\includegraphics[trim=0.0cm 0cm 0.0cm 0cm, clip=true, scale=.50]{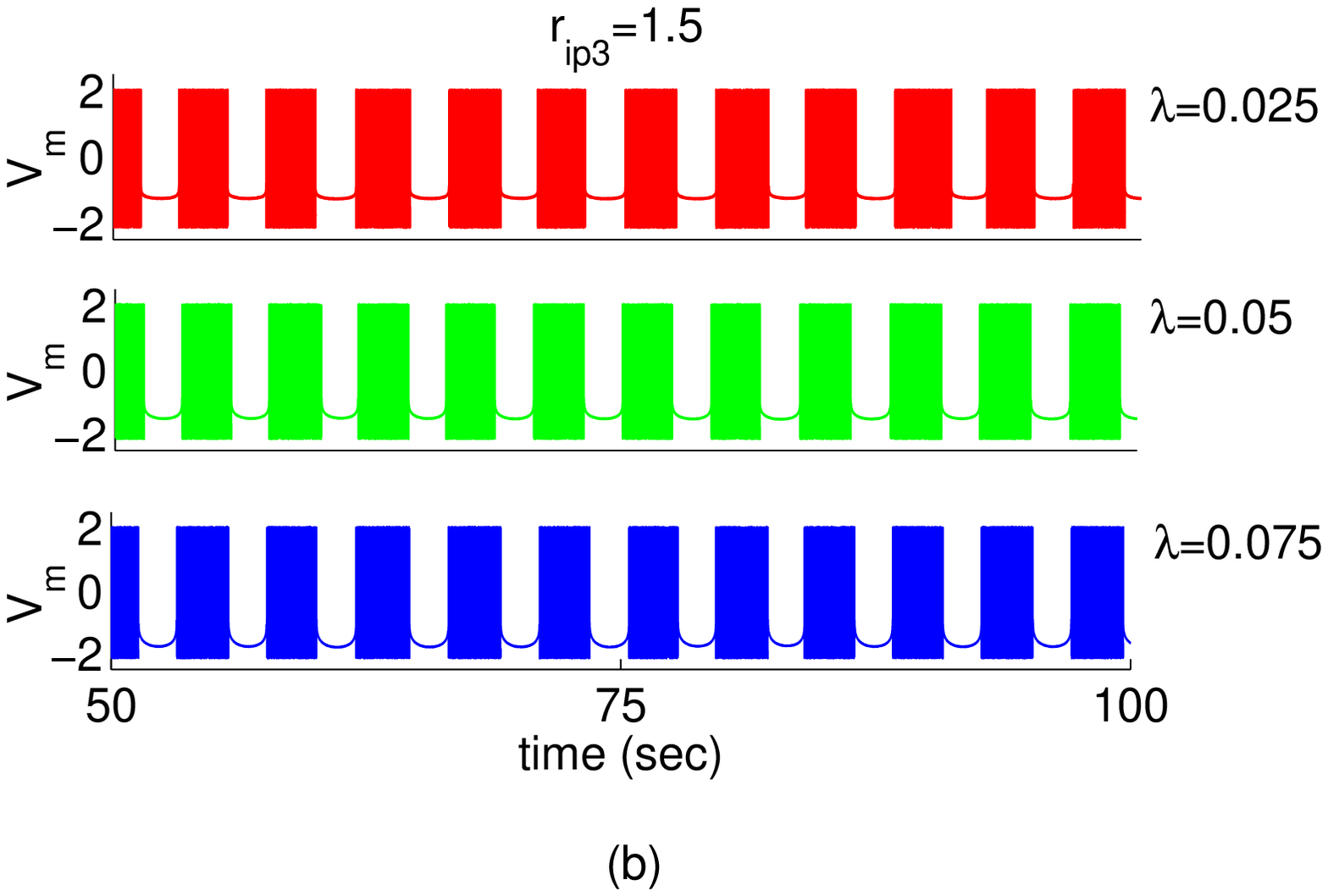}
		\includegraphics[trim=0.0cm 0cm 0.0cm 0cm, clip=true, scale=.50]{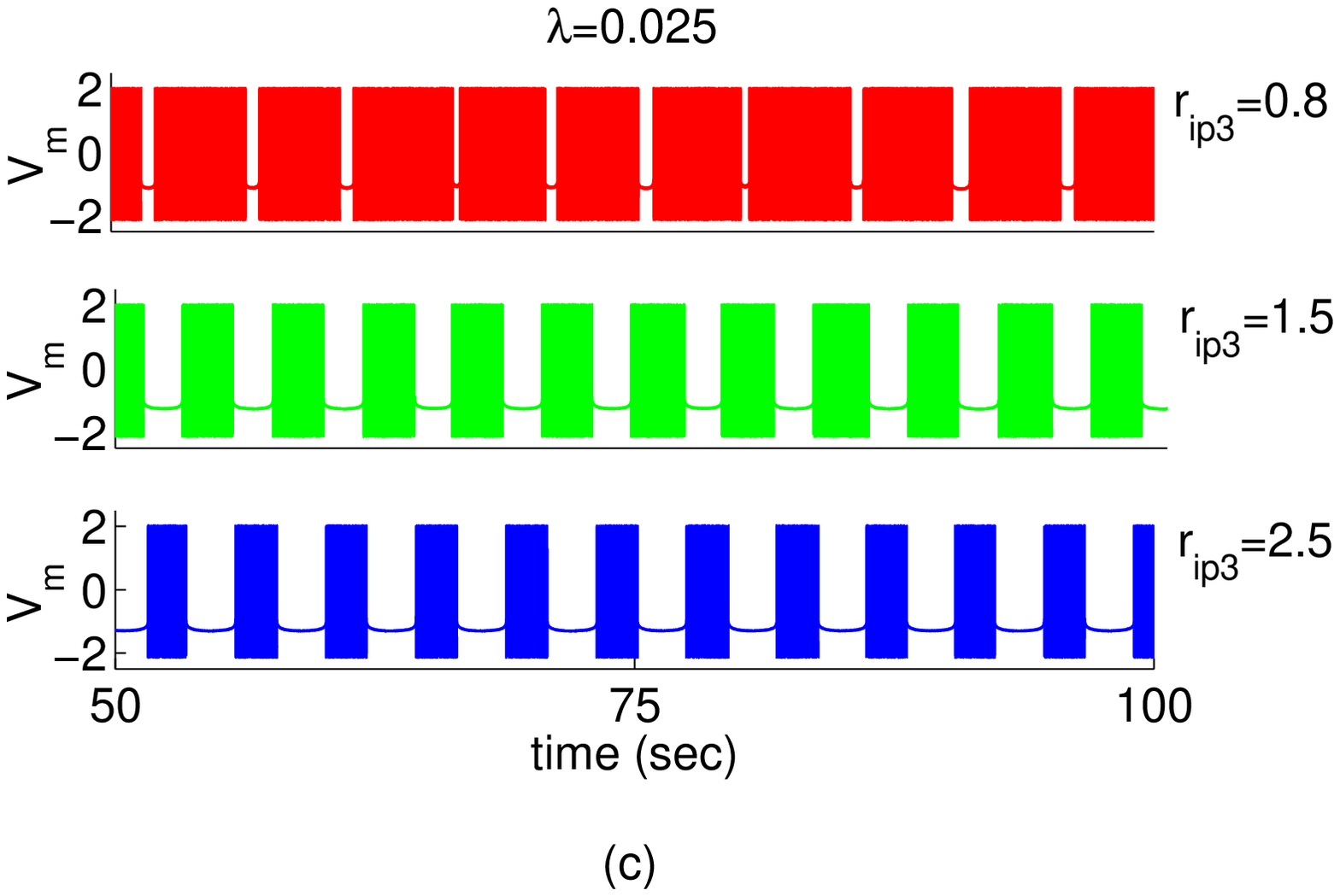}
		\includegraphics[trim=0.0cm 0cm 0.0cm 0cm, clip=true, scale=.26]{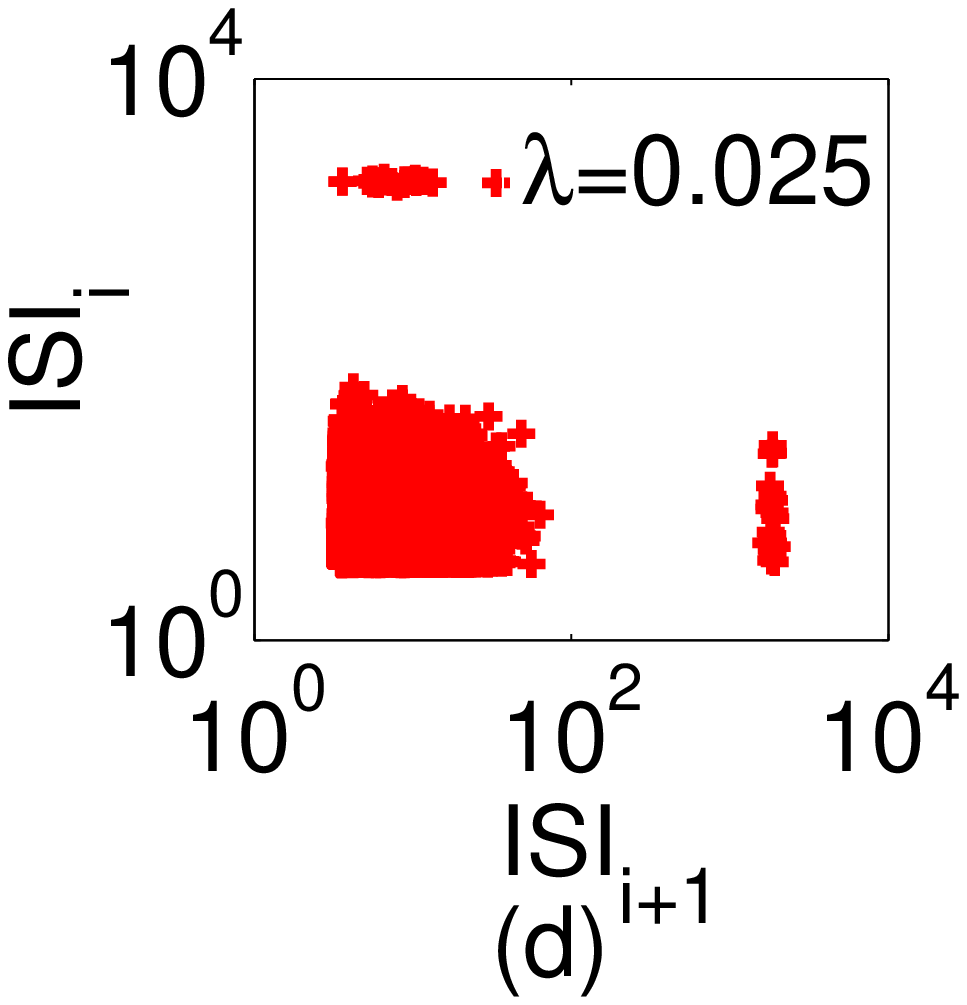}
		\includegraphics[trim=0.0cm 0cm 0.0cm 0cm, clip=true, scale=.26]{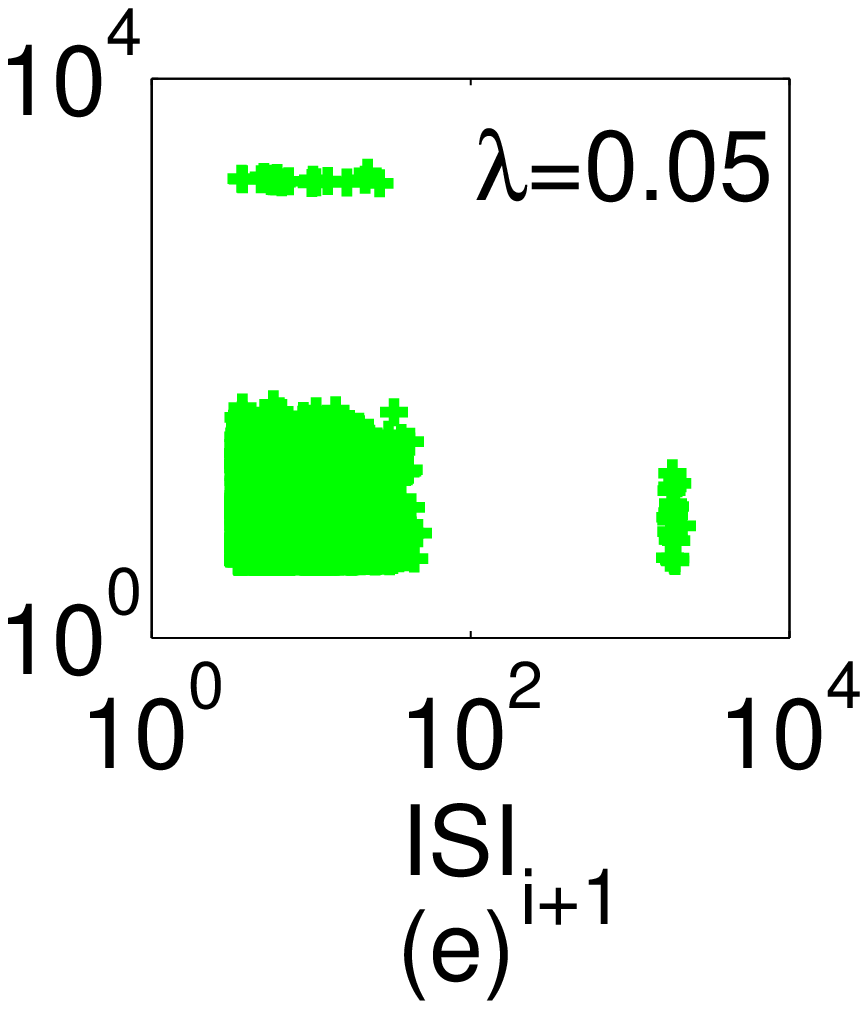}
		\includegraphics[trim=0.0cm 0cm 0.0cm 0cm, clip=true, scale=.26]{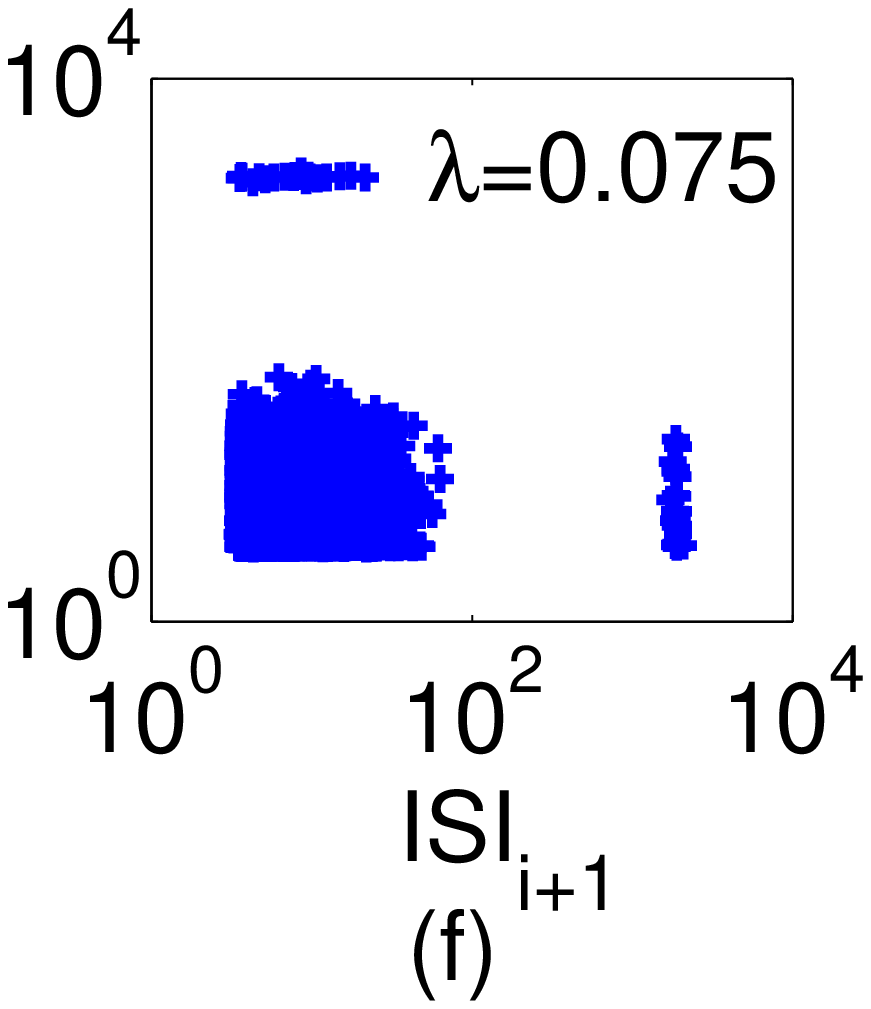}
		\includegraphics[trim=0.0cm 0cm 0.0cm 0cm, clip=true, scale=.26]{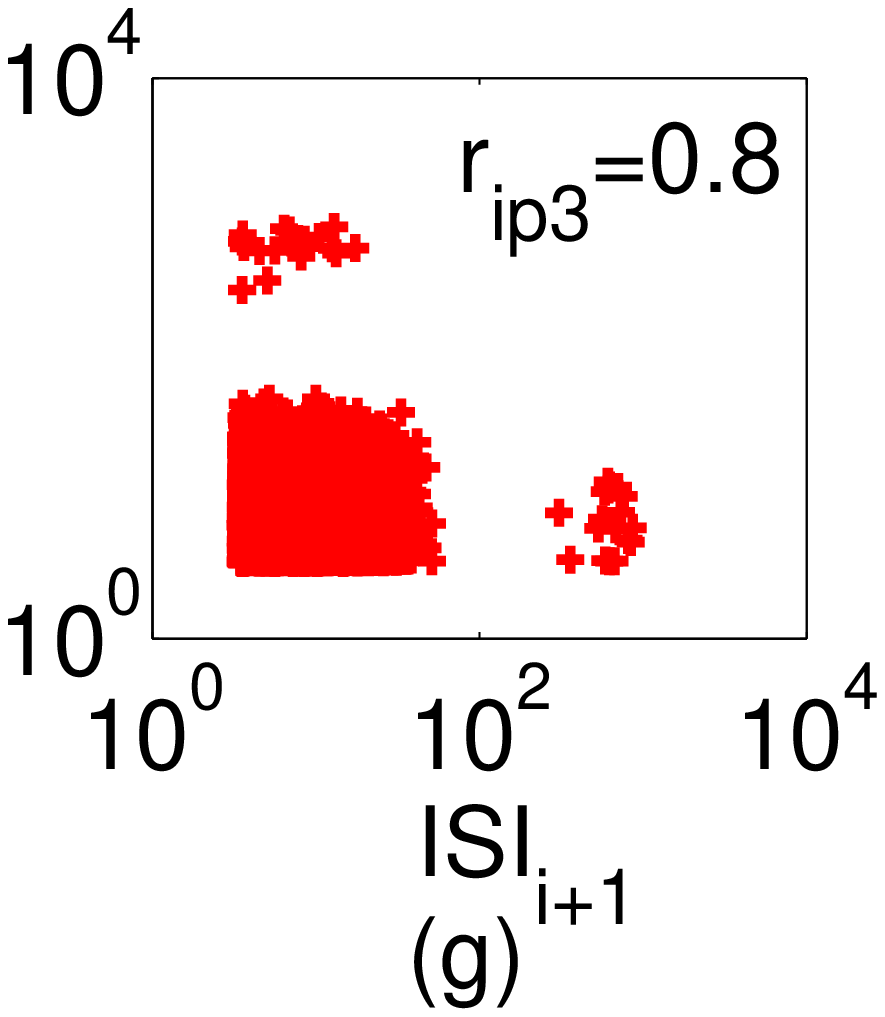}
		\includegraphics[trim=0.0cm 0cm 0.0cm 0cm, clip=true, scale=.26]{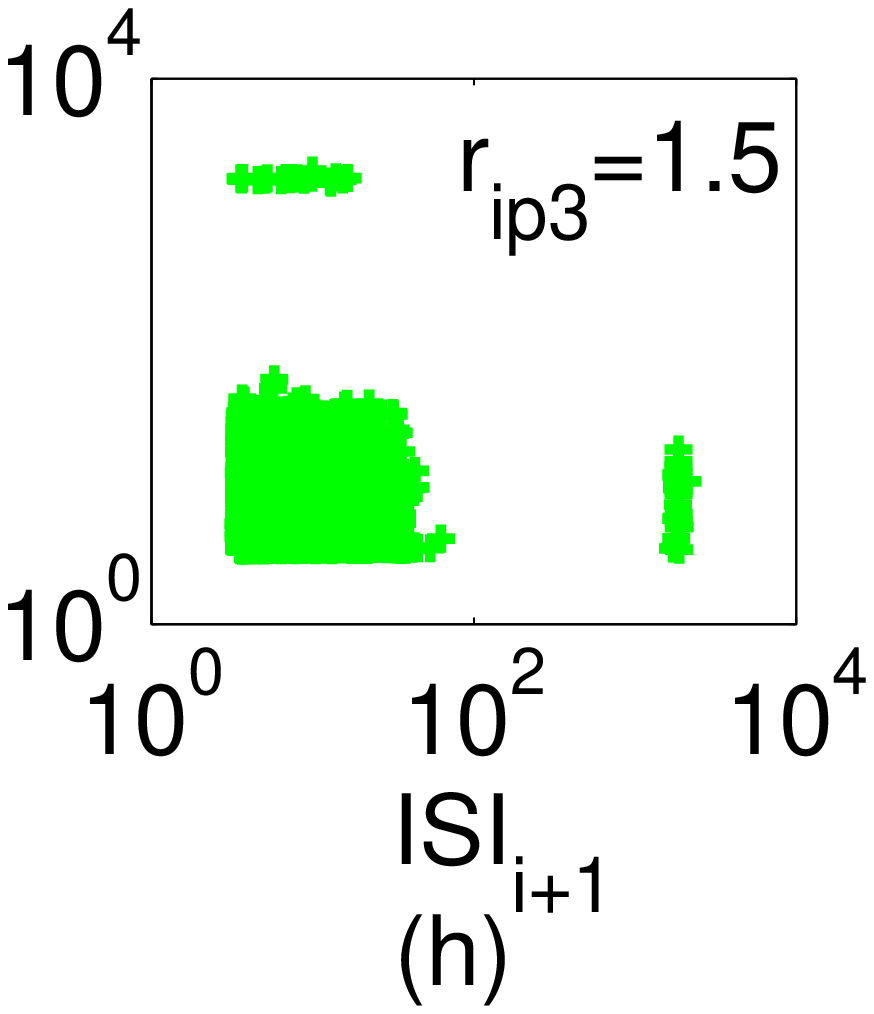}
		\includegraphics[trim=0.0cm 0cm 0.0cm 0cm, clip=true, scale=.26]{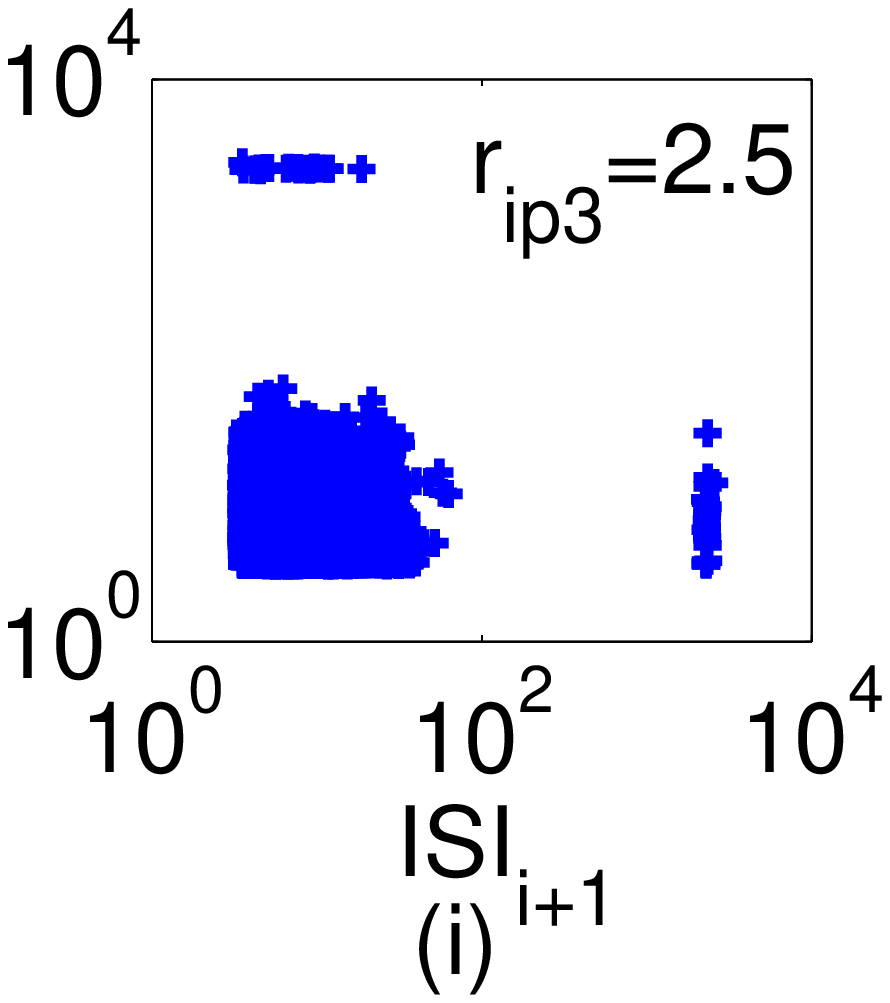}
		\caption{(a) Schematic illustration of conceptual explanation of statistical parameters $BD$, $IBI$, $ISI$. (b) Voltage traces of the neuron for different values of $\lambda$ when $r_{ip3}=$1.5. (c) Voltage traces of the neuron for different values of $r_{ip3}$ when $\lambda$ is fixed to 0.025. (d-f) 2D $ISI$ histograms computed through very long simulation time for different values of $\lambda$. (g-i) The same analyses as in panels (d-f) for different values of $r_{ip3}$}.
	\end{figure*}	
	
	So far, the intracellular $Ca^{2+}$ dynamics that underlies the generation of $I_{astro}$ has been considered with a deterministic approach where many local and global processes taking place in the cellular environment are ignored. This approach simply models the activation of the astrocyte current. However, metabolic processes are noisy due to many external and internal factors. For instance, the astrocytic second messenger $IP3$ sensitizes $IP3$ receptor channels ($IP3Rs$) mediating $Ca^{2+}$ release from the endoplasmic reticulum. $IP3Rs$ are typically distributed in clusters with several channels on the $ER$ membrane of the astrocyte. The noise caused by the random opening and closing of these channels is expressed as $Ca^{2+}$ channel noise of the astrocyte which is already included in our model (see Eq. (5)). Accordingly, if the clusters comprise a few channels,  $Ca^{2+}$ release occurs stochastically, thus generating noise, and the level of noise is dependent on the  size of the cluster, i.e., the number of $IP3Rs$ denoted by $Nt$. Thus, we utilize a stochastic $LR$ model combined with $Langevin$ approach  for $Ca^{2+}$ release from $ER$ that has proven accurate for a variety of cluster sizes  \cite {nadkarni2007}. In Fig. 7, we show how astrocytic noise, which is controlled by the cluster of $Nt$ channels,  affects the resulting behavior of $CV$ as a function of neuronal noise level $D$ in our model setup. As seen in the figure, regardless of the $Nt$ values, the $CV$ curves exhibit a bell-shaped  dependence on neuronal noise, indicating the presence of an optimal $D$ for maximum firing irregularity. Moreover, $CV$ values increase as the level of the astrocytic noise increases in the system (decreasing $Nt$). It can be observed that the decrease in $Nt$ decreases the optimal $D$  corresponding to the peak $CV_{max}$. This indicates that it requires less spike inputs from the neuron that is controlled by $D$ to obtain the same firing irregularity level in the presence of astrocytic noise in the neuron-astrocyte coupled model. Finally, for too high neuronal noise levels, the system is dominated by the neuronal noise; astrocytic noise then becomes insignificant and intense firing activity of the neuron due to high noise level causes the $CV$ to decrease. These findings indicate that, for fixed parameter sets ($a_{e}$, $\lambda$ and $r_{ip3}$), the density of channels influences the neuronal dynamics distinctly in the coupled system.

	\subsection {Bursty Regime with Astrocyte}
	
In recent decades, many researchers have suggested that the astrocyte  provides an important physiological mechanism to enable bursty firing activity \cite{ Morquette2015, Deemyad2018}. We have already showed the emergence of such a bursty mode at the proper coupling strength $\lambda$ (see Fig. 1). In this section, based on statistical analyses, we investigate whether it is possible to control the frequency and size of bursty firings with the dynamics of the astrocyte. To do so, we first describe some statistical notations to characterize the bursty regime as illustrated in Fig. 8a. The $IBI$ (interburst interval) is the period between two bursts, the $BD$ (burst duration) is the time period between the beginning and the end of each burst period, and finally, the $ISI_{i}$ denotes the \textit{i-th} interspike-interval \cite{womack2004, fardet2018}. After providing the necessary descriptive information, we now show the effect of $\lambda$ and $r_{ip3}$ on $IBI$ and $BD$ in Fig. 8b and Fig. 8c respectively, by monitoring the membrane potential traces. 

The $V_{m}$ voltage traces of the neuron in the left panel, which has the same characteristic even though the $\lambda$ increases, clearly shows that it does not change the frequency and size of the burst window. On the other hand, with the increased values of $r_{ip3}$, it is seen that $IBI$ increases and $BD$ decreases. This indicates that $r_{ip3}$ can control the burst characteristic. Another point draws attention: the first $IBI$ period occurs earlier as $r_{ip3}$ increases because the astrocyte is activated in a much shorter time with larger $IP3$ production rate. 

To check whether this bursty activity mode is not a transient behavior, 2D $ISI$ histograms are obtained for very long periods of simulation time with the same $\lambda$ and $r_{ip3}$ values considered in upper panels. These charts shown in panels (d-i) present detailed information about whether the neuron exhibits bursty firing or regular spiking behavior without observing the membrane traces. Namely, the large clusters located at relatively small $ISI_{i}$ and $ISI_{i+1}$ values indicate the firing activity during the burst, while the other two small clusters in histograms denote the starting and the completion of the burst periods. In Figs. 8(d,e,f), it is clear that the size and the scattering of clusters do not change very much with $\lambda$. However, as seen in Figs. 8(g,h,i), small clusters get less noisy as $r_{ip3}$ increases. These observations from Fig.8 indicate that the coupling between the astrocyte and the neuron is responsible for the emergence of the bursty firing mode while it has no significant influence on bursting statistics (i.e. $IBI$). It can also be inferred that $r_{ip3}$ might be a critical player in the regulation of the bursty firing mode statistics.

To support these findings, we systematically investigate the statistical changes in bursty regime through average and standard deviation of $BD$ and $IBI$ measures as functions of $\lambda$ and $r_{ip3}$ for three different neuronal noise levels. Fig. 9 features the obtained results. It is seen that the mean ($<BD>$) and standard deviation ($\sigma_{BD}$) of $BD$ and $IBI$ are approximately constant as $\lambda$ varies for any given levels of noise (see panels (a) and (c)). This indicates that our finding from Fig. 8(d-f), i.e. that is $\lambda$ has no role on burst statistics, is robust to neuronal noise. The most striking point here is that neuronal noise acts as a control parameter for the lifetimes of the bursts in such a neuron-astrocyte crosstalk environment. In other words, with increased values of $D$, the emergent intense neuronal discharges result in stable (i.e. periodic and almost equal amplitude) astrocytic currents in time due to the rapid  $Ca^{2+}$ accumulation in the cytosol of the astrocyte. Thus, small $<BD>$ and large $<IBI>$ occur as $D$ increases. Additionally,  since the stable astrocytic current pulses trigger regular and consistent burst activity for large $D$, $\sigma_{BD}$ and $\sigma_{IBI}$ become quite small. On the other hand, panels (b) and (d) illustrate that $r_{ip3}$ has a significant role on $BD$ and $IBI$ statistics. The $<BD>$ decreases with $r_{ip3}$ at all considered $D$ levels, while the $<IBI>$ shows the opposite direction of change from the $<BD>$ (as expected). We know from previous analyses that as $r_{ip3}$ increases, astrocyte current emerges with greater amplitude and lifetime. Thus, $<BD>$ decreases for larger values of $r_{ip3}$, which results in longer cessation of firing activity during the periods of time when the astrocyte is active. It is also clear from panels (b) and (d) that the influence of neuronal noise ($D$) on burst statistics under the variation of $r_{ip3}$ exhibits similar behavior as that observed in the case of $\lambda$ variation (panels (a) and (c)).
	\begin{figure}[htp]
		\centerline{\includegraphics[trim=0.0cm 0cm 0.0cm 0cm, clip=true, scale=.35]{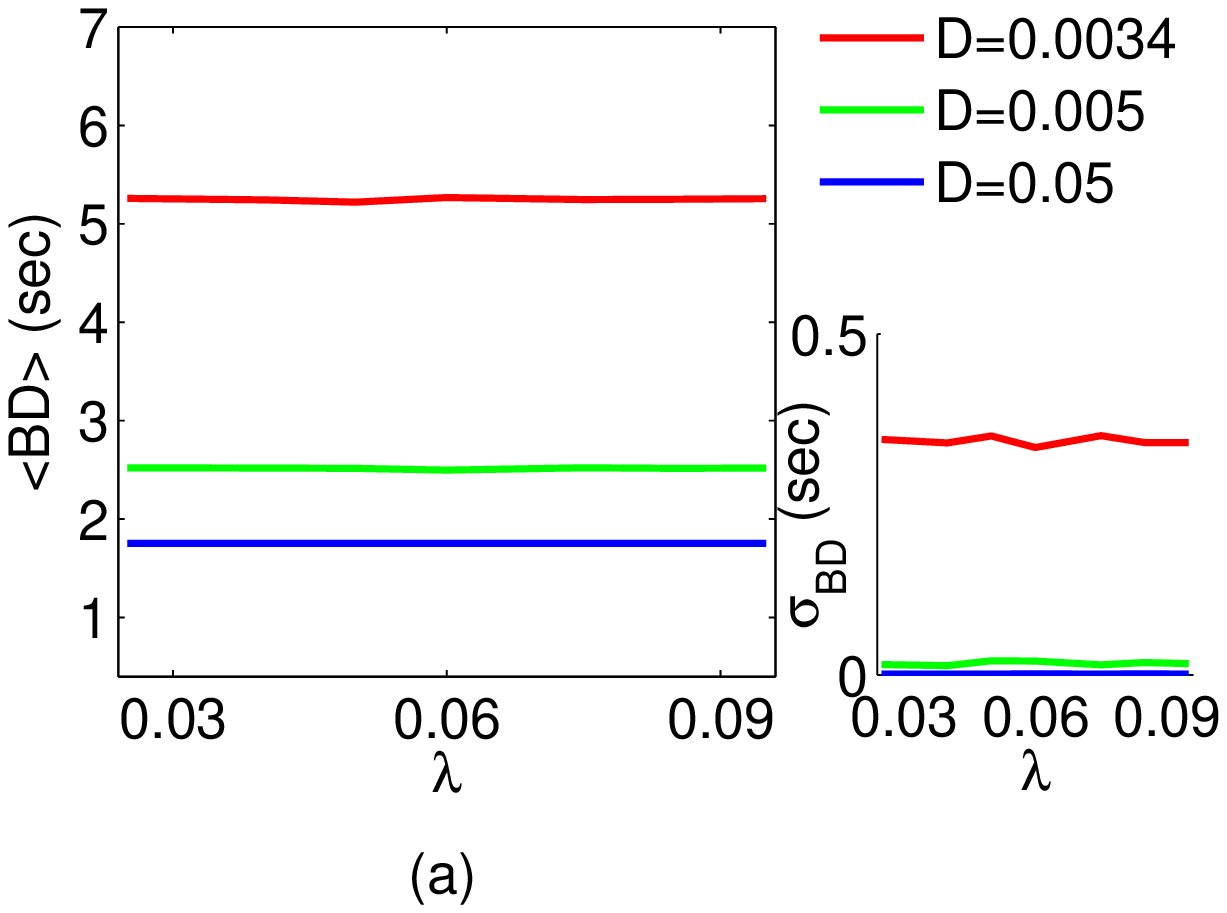}
		\includegraphics[trim=0.0cm 0cm 0.0cm 0cm, clip=true, scale=.35]{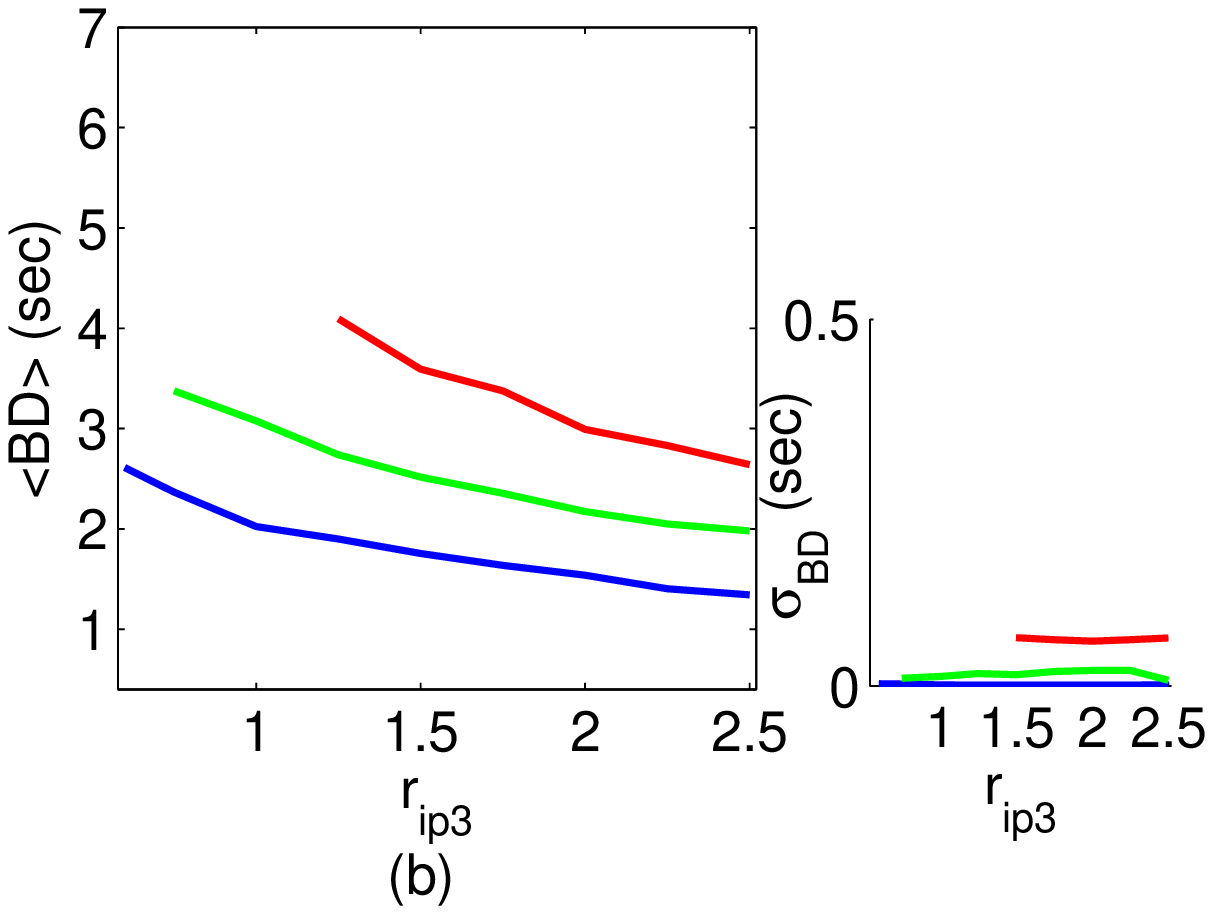}}
	\centerline{
		\includegraphics[trim=0.0cm 0cm 0.0cm 0cm, clip=true, scale=.35]{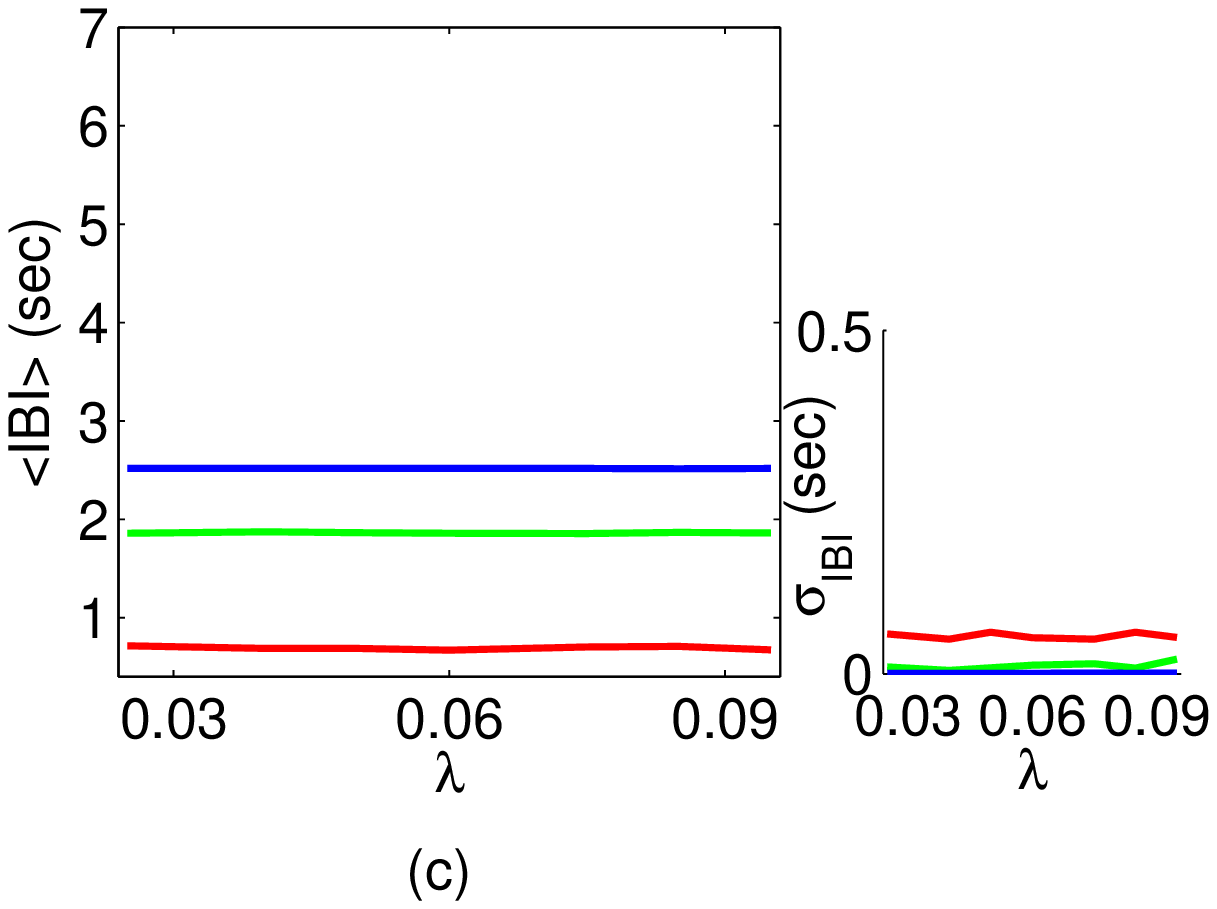}
		\includegraphics[trim=0.0cm 0cm 0.0cm 0cm, clip=true, scale=.35]{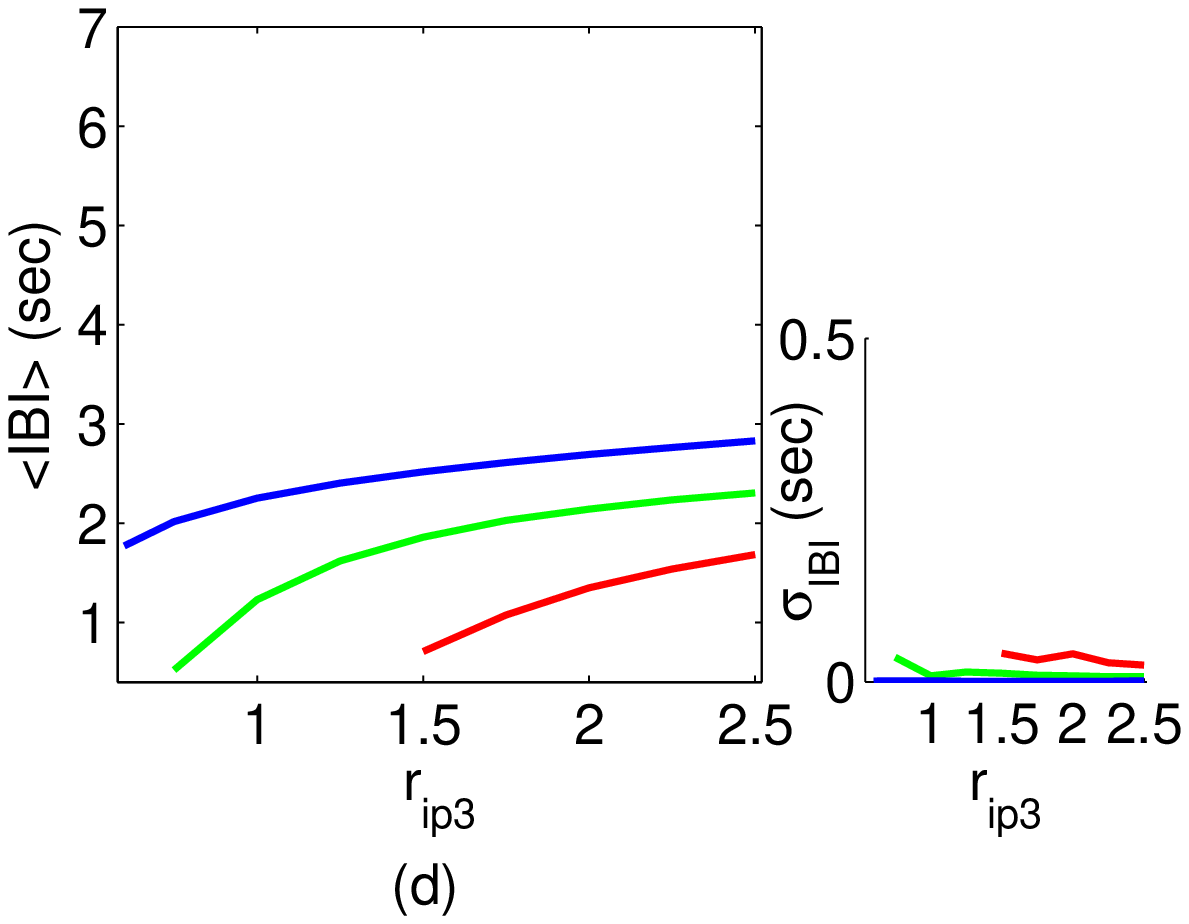}}
		\caption{The variation of the average and standard deviation of burst duration and interburst interval as a function of  $\lambda$ and $r_{ip3}$ for different values of $D$. (a) Variation of the mean $<BD>$ and standard deviation $\sigma_{BD}$ of burst duration as a function of $\lambda$. (b) The same analyses as in panel (a) for $r_{ip3}$. (c) The mean $<IBI>$ and standard deviation  $\sigma_{IBI}$ of interburst-interval versus $\lambda$. (d) The same analyses as in panel (c) for $r_{ip3}$. Here, the excitability control parameter of the neuron is fixed to $a_{e}=1.01$.}
	\end{figure}

		\begin{figure}[htp]
		\centerline{\includegraphics[trim=0.0cm 0cm 0.0cm 0cm, clip=true, scale=.4]{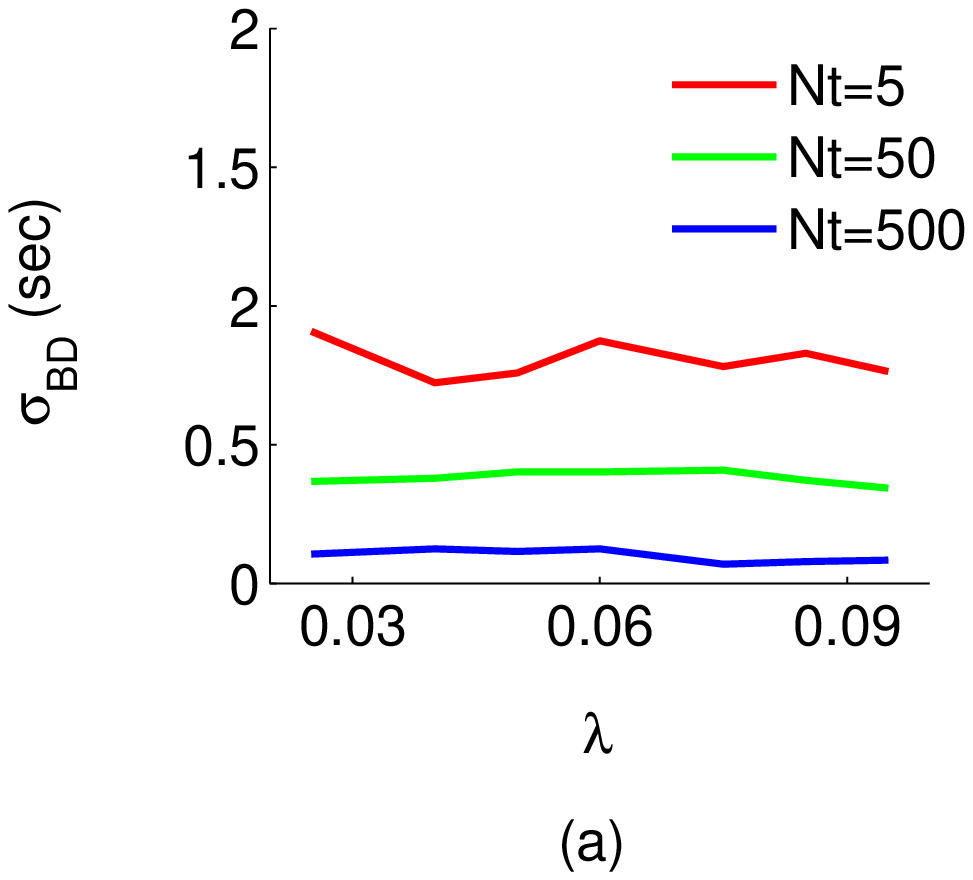}
		\includegraphics[trim=0.0cm 0cm 0.0cm 0cm, clip=true, scale=.4]{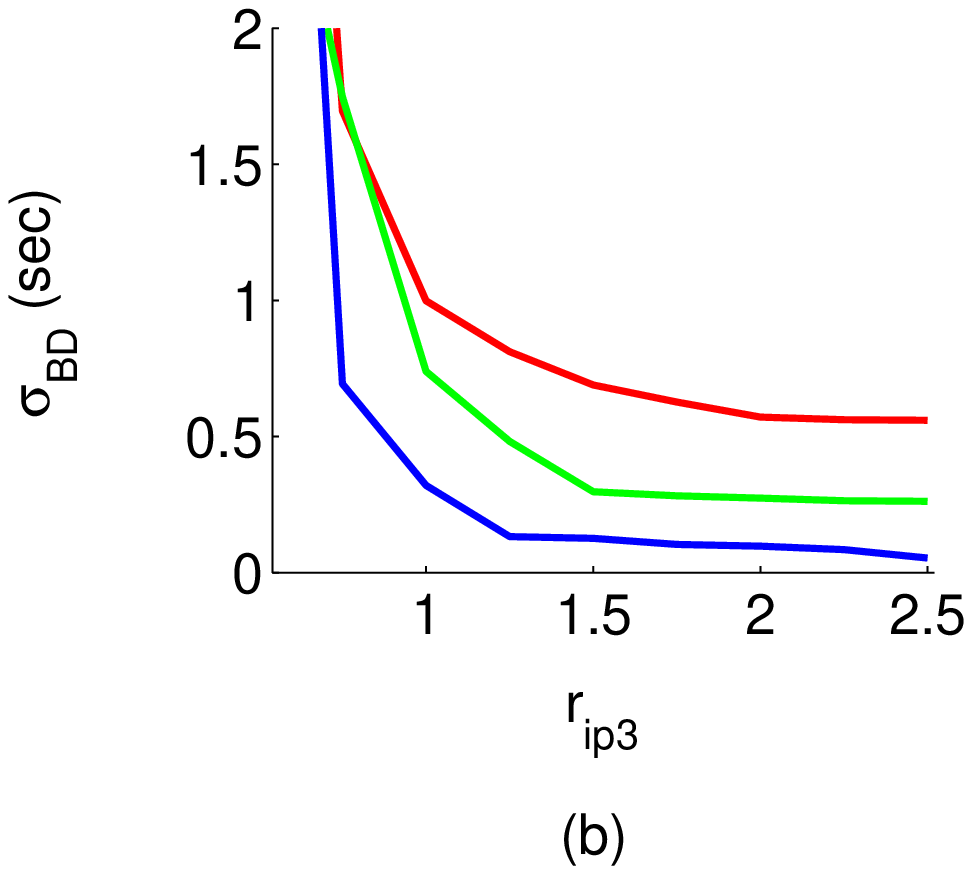}}
	 \centerline{
		\includegraphics[trim=0.0cm 0cm 0.0cm 0cm, clip=true, scale=.4]{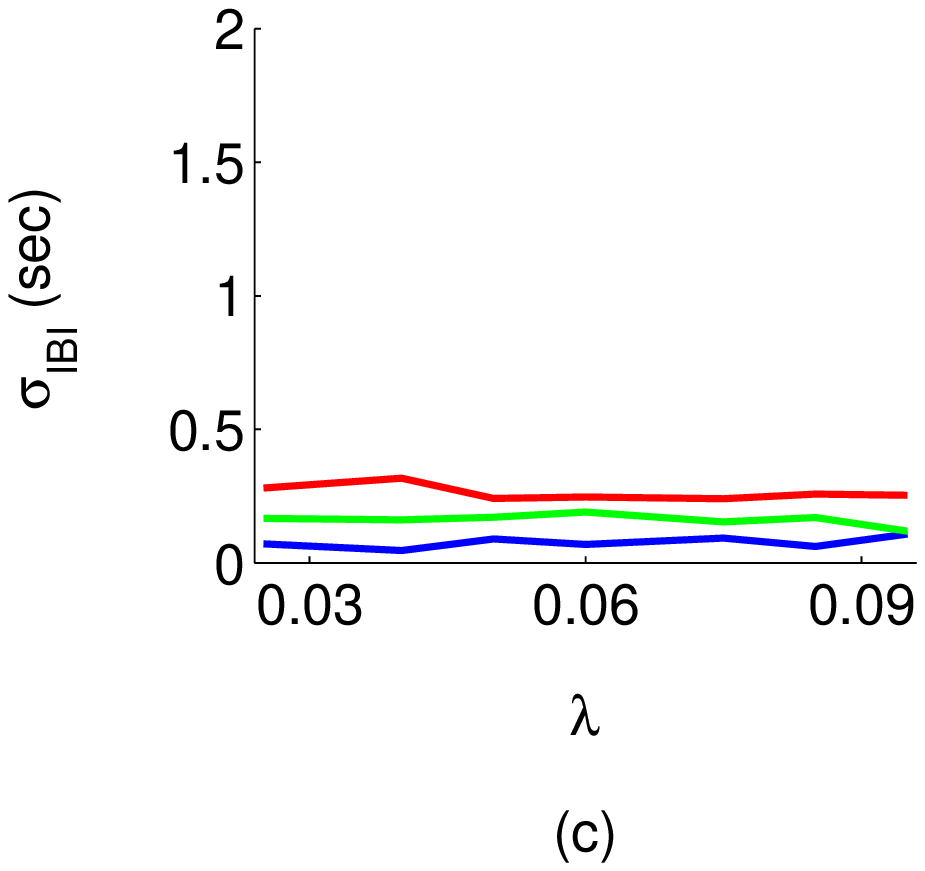}
		\includegraphics[trim=0.0cm 0cm 0.0cm 0cm, clip=true, scale=.4]{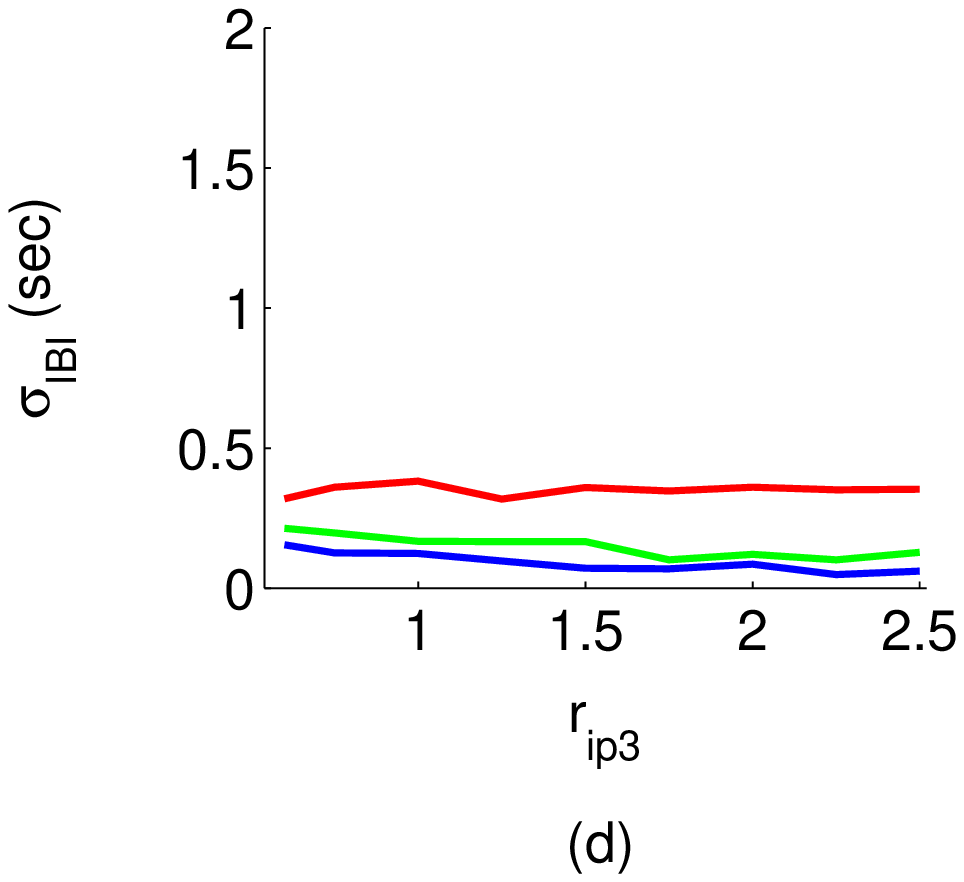}}
		\caption{The variation of $\sigma_{BD}$ and $\sigma_{IBI}$ as functions of  $\lambda$ and $r_{ip3}$ for different values of $Nt$. (a) The dependence of $\sigma_{BD}$ on $\lambda$ and (b) on $r_{ip3}$. (c) Standard deviation of interburst-interval $\sigma_{IBI}$ versus $\lambda$. (d) The same analyses as in panel (c) for $r_{ip3}$. The presented results are obtained for a fixed parameter set, i.e. $D=0.005$, $a_{e}=1.01$.}
	\end{figure}

	\begin{figure}[htp]
		\includegraphics[trim=0.0cm 0cm 0.0cm 0cm, clip=true, scale=0.6]{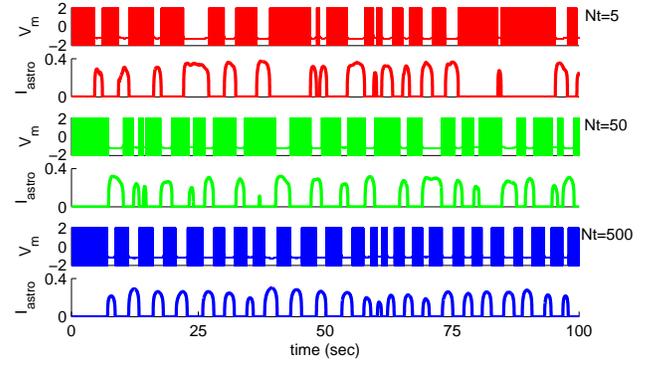}
		\caption{Time variation of the membrane potential of the neuron and the feedback current $I_{astro}$ for different values of the number of $IP3Rs$. For each $Nt$ values, we plot the voltage traces of the neuron (top panel)  and $I_{astro}$  feedback currents (bottom panel) by using the same color code for fixed $r_{ip3}=1.5$ and $\lambda=0.025$. The other parameters of the model are $a_{e}=1.01$ and $D=0.005$.}
	\end{figure}

	Finally, we investigate how the intrinsic noise stemming from random nature of $IP3R$ channels affects the bursty firing mode. To do so, we compute the standard deviation of $IBI$ and $BD$ by fixing the neuronal noise level to $D=0.005$ under the variations of $\lambda$ and $r_{ip3}$. Results are presented in Fig. 10. It is seen that as the astrocytic noise increases ($Nt$ decreases), $\sigma_{IBI}$ and $\sigma_{BD}$ increase. This finding is more pronounced for $\sigma_{BD}$ because the $BD$ windows are larger than the $IBI$ windows. On the other hand, we still observe that $\lambda$ has no role under the variation of $Nt$ (see panel a and c). However, $r_{ip3}$ can be expressed as a control parameter that allows modification of burst patterns. It is seen from panel (b) that $\sigma_{BD}$ first gradually decreases, and then, saturates to an approximately fixed level with the increase of $r_{ip3}$. When the effects of $Nt$ and $D$ are compared on burst mode firing, it can be said that although the trends exhibit similarity, the standard deviation of burst statistics under the variation of $Nt$ is much larger than for the case of $D$ (around 100 times). The reason of such an extraordinary distinction arises from the fact that $Nt$ and $D$ shape the characteristics of the astrocytic current. Recall that the increase of $D$ results in a more stable $I_{astro}$ both in time and amplitude. However, as can be observed from Fig. 11 where we illustrate $I_{astro}$ and the corresponding $V_{m}$ voltage traces of the membrane (with same color-codes) for three different values of $Nt$, $I_{astro}$ becomes unstable with the increase of astrocytic noise (decrease of $Nt$). Namely, it is obvious that the width and position of $I_{astro}$ pulses in time significantly becomes random as $Nt$ decreases, which results in the emergence of both long and short burst periods under the same biophysical conditions (see voltage traces in Fig.11). Thus, $\sigma_{IBI}$ and $\sigma_{BD}$ reach larger values as the astrocytic noise increases compared to the case of the increase in neuronal noise.

	\section{Conclusion}

	In this work, we study the effect of an astrocyte on the spontaneous firing activity of a neuron that is initially subject to noise arising from biophysical conditions. To do so, we used a bipartite model that includes a neuron and an astrocyte and explored the parameter space that may induce intriguing neuron firing behaviors which are not present without astrocytes (isolated neurons). First, through an astrocytic parameter space analysis, it was shown that, in addition to the noise-induced spontaneous firing regime, a new bursty firing mode may emerge that does not exist in an isolated FHN neuron dynamics. We explained this astrocyte induced new firing mode through analysis of the feedback current $I_{astro}$ introduced into the neuron dynamics. When there exists a crosstalk between these two different cells, the  membrane voltage of the neuron acts on the intrinsic calcium dynamics of the astrocyte, which in turn affects the distance to threshold of the excitability level of the neuron through $I_{astro}$. Depending on the waveform (width and amplitude) and the frequency of the feedback current provided by astrocyte, we showed that this interaction may result in either a spontaneous or bursty firing mode. 
	
	In order to observe what would be the impact of astrocytes on these firing regimes, we restricted our analyses to the most significant features of each individual firing behavior. We first investigated the coherence of noise-induced spontaneous firing behavior of a single neuron that is coupled to an astrocyte. Compared to the isolated case, our findings revealed that the inherent dynamics of an astrocyte can induce more irregular firing patterns of spontaneous spiking activity for a particular range of noise intensity. It was also shown that the astrocytic parameters $\lambda$ and $r_{ip3}$ provide a control mechanism for the level of such irregularity. We then extended our analyses by adding an additional noise source into our system, stemming from stochastic nature of $IP3$ receptor channels that control $Ca_{2+}$ release inside the astrocyte. It was found that the resonance behavior of $CV$ as a function of neuronal noise remains in such a coupled system.  Further, the astrocytic calcium noise enhances this effect by reducing  the required neuronal noise intensity for the same irregularity level.

	 Secondly, we studied whether it is possible to control the basic features of the bursty firing mode with astrocyte dynamics. By analyzing the width and regularity of bursty periods, we concluded that the coupling strength $\lambda$ between astrocyte and neuron, which has a key role in determining the firing regime, does not play a significant role in modulation of bursty behavior, but the $IP3$ production rate in astrocyte $r_{ip3}$ has the ability to control all considered features of burst firing mode.  Moreover, our findings suggest that, in such a neuron-astrocyte crosstalk environment,  neuronal and astrocytic noise act as a control parameter for determining periodic (the same burst width) and stochastic (random burst width) characteristic of $BD$ and $IBI$ statistics, respectively. 

Finally, we would like to emphasize that our study provides an alternative theoretical framework indicating the power of astrocytes in shaping the neural activity. However, since the system under study in this work includes a single neuron-astrocyte pair, we cannot generalize our findings to a population scale. Thus, a possible extension of our study could be investigating the dynamics of spontaneous and burst firing modes in networks of neurons and astrocytes, which have been mostly studied so far in only neuron populations \cite{ma2017, bera2019} by considering various biophysical circumstances such as network topology and the type of gliotransmitters.

\begin{acknowledgments}
M.U. acknowledges the support of the Scientific and Technological Research Council of Turkey (TUBITAK) BIDEB-2219 International Postdoctoral Research Fellowship Program for his research stay in Ottawa, Canada.
\end{acknowledgments}

\section*{AUTHOR DECLARATIONS}

\subsection*{Conflict of Interest}

The authors have no conflicts to disclose.

\section*{Data Availability Statement}

Data sharing is not applicable to this article as no new data were created or analyzed in this study.

\end{document}